\documentclass[journal]{IEEEtran}
\usepackage{subfig}
\usepackage[pdftex]{graphicx}
\usepackage{tikz}
\usetikzlibrary{spy}
\usepackage[colorlinks=true,linkcolor=blue,citecolor=blue]{hyperref}
\usepackage{url}
\usepackage{epstopdf}
\usepackage{amsmath,amssymb,amsfonts}
\usepackage{mathrsfs}
\usepackage{float}
\usepackage{cleveref}
\usepackage{booktabs}
  
\usepackage{algorithm}
\usepackage{algpseudocode}

\newcommand*\Let[2]{\State #1 $\gets$ #2}

\usepackage{acro}
\usepackage{paperacronyms}
\acuse{GPU,CPU}

\newcommand{\Real}{\ensuremath{\mathbb{R}}}

\newcommand{\RecSpace}{\ensuremath{X}}
\newcommand{\DataSpace}{\ensuremath{Y}}

\newcommand{\approxInv}[1]{#1^{\dagger}}
\DeclareMathOperator{\ForwardOp}{\ensuremath{\mathcal{A}}}
\DeclareMathOperator{\ForwardOpPseudoInv}{\ensuremath{\approxInv{\ForwardOp}}}
\DeclareMathOperator{\ForwardOpInvLearned}{\ensuremath{\approxInv{\mathcal{A}}_{\param}}}

\DeclareMathOperator{\ForwardOpAdj}{\ensuremath{\ForwardOp^*}}

\DeclareMathOperator*{\argmin}{arg\,min}

\DeclareMathOperator{\Loss}{\ensuremath{L}}

\DeclareMathOperator{\Network}{\ensuremath{\Lambda}}
\newcommand{\param}{\ensuremath{\theta}}

\newcommand{\signal}{\ensuremath{f}}
\newcommand{\signaltrue}{\signal^{*}}
\newcommand{\signalest}{\hat{\signal}}
\newcommand{\data}{\ensuremath{g}}
\newcommand{\noise}{\ensuremath{\delta\data}}

\DeclareMathOperator{\ProjectScale}{\pi}
\DeclareMathOperator{\UpSample}{\tau}

\newcommand{\rev}[1]{{#1}}

\title{Multi-Scale Learned Iterative Reconstruction}
\author{Andreas Hauptmann, Jonas Adler, Simon Arridge, and Ozan \"Oktem
}
\begin{document}

\author{Andreas Hauptmann,
	 	Jonas Adler,
  	 	Simon Arridge, 
  	 	and Ozan \"Oktem
\thanks{This work was partially supported by the Academy of Finland Project 312123 (Finnish Centre of Excellence in Inverse Modelling and Imaging, 2018--2025) and Project 334817, as well as British Heart Foundation grant NH/18/1/33511, EPSRC grant EP/M020533/1, and EPSRC-Wellcome grant WT101957.
The authors would also like to acknowledge the Swedish Foundation for Strategic Research grants \emph{Low complexity reconstruction for medicine} (AM13-0049) and \emph{3D reconstruction with simulated image formation models} (ID14-0055). Funding has also been provided by Elekta (Stockholm, Sweden).}%
\thanks{AH and JA contributed equally}%
\thanks{A. Hauptmann is with the Research Unit of Mathematical Sciences; University of Oulu, Oulu, Finland and with the Department of Computer Science; University College London, London, United Kingdom.}%
\thanks{J. Adler did this work at Elekta, Stockholm, Sweden and KTH -- Royal Institute of Technology, Stockolm, Sweden. He is currently with DeepMind, London, UK.}%
\thanks{S. Arridge is with the Department of Computer Science; University College London, London, United Kingdom.}%
\thanks{O. \"Oktem is with is with the Department of Mathematics, KTH -- Royal Institute of Technology, Stockholm, Sweden.} }%

		\maketitle
	
	
\begin{abstract}
Model-based learned iterative reconstruction methods have recently been shown to outperform classical reconstruction algorithms. Applicability of these methods to large scale inverse problems is however limited by the available memory for training and extensive training times\rev{, the latter} due to computationally expensive forward models. As a possible solution to these restrictions we propose a multi-scale learned iterative reconstruction scheme that computes iterates on discretisations of increasing resolution. This procedure does not only reduce memory requirements, it also considerably speeds up reconstruction and training times, but most importantly is scalable to large scale inverse problems with non-trivial forward operators, 
such as those that arise in many 3D tomographic applications. In particular, we propose a hybrid network that combines the multi-scale iterative approach with a particularly expressive network  architecture which in combination exhibits excellent scalability in 3D.

Applicability of the algorithm is demonstrated for 3D cone beam computed tomography from real measurement data of an organic phantom. Additionally, we examine scalability and reconstruction quality in comparison to established learned reconstruction methods in two dimensions for low dose computed tomography on human phantoms.
\end{abstract}

	
	

\section{Introduction}
\Ac{CT} is an imaging technology where the interior anatomy of a subject is computed from a series of X-ray radiographs acquired by radiating the subject from different directions. \Ac{CT} has had a profound impact on medical practice and it is now an indispensable technology in a wide spectrum of clinical and industrial applications. It has also been essential for advancing our understanding of disease in medical research.

\rev{\ac{CT} imaging is however associated with risks}, especially when \rev{it is} used for screening. \Ac{CT} relies on repeatedly exposing a patient to \rev{ionising radiation of X-rays} and hence there is an ongoing effort to minimise the total dose delivered to a patient during a \ac{CT} scan. For that purpose, low dose \ac{CT} protocols \rev{can} be employed \rev{where fewer} X-ray photons are measured\rev{, which consequently reduces the signal-to-noise ratio in acquired data}. \rev{Widely used reconstruction techniques in clinical practice, such as filtered backprojection,} are based on sampling theory and as such are not properly adapted to account for the statistical characteristics of measured data with high noise level. Hence, applying these schemes on low-dose \ac{CT} data will produce sub-optimal images which consequently prevents low dose protocols from being widely adapted. 
Furthermore, in industrial and scientific applications which often utilise $\mu$\acs{CT} systems, reconstructions are typically computed by the \ac{FDK} algorithm \cite{feldkamp1984practical} used for cone beam \ac{CBCT} measurements.
Here the same requirement of many angles applies, but additionally reconstructions often exhibit cone beam artefacts due to the measurement geometry. Accurate measurement procedures to overcome these issues can be highly time consuming and effectively limit experimental capacity, hence there is a need for advanced and computationally efficient reconstructions algorithms from few angle measurements.

Over the years, a wide range of reconstruction methods have been developed that better account for the aforementioned statistical properties in few angle and low-dose \ac{CT} scans. Among these, the most powerful and flexible have been variational model-based methods \cite{Stiller:2018aa,Geyer:2015aa,Sidky2012convex}. These offer a plug-and-play architecture for reconstruction where a user provides a model for how data is generated in absence of noise (forward operator), a statistical model for noise in data, and a prior model for desired reconstructions. The forward operator together with the statistical model for data ensures consistency against measured data, whereas the prior mainly prevents over-fitting by penalising images that have `irregular' behaviour. \rev{These variational methods can sometimes be interpreted as computing the most likely solution (maximum a posteriori estimate) \cite[sec.~3.3.2]{ActaReview}, see also \cite{Kaipio2004a,Siltanen2003statistical,Kolehmainen2003statistical}.
Variational model-based reconstruction is, however,
computationally demanding since it involves solving a large-scale optimisation problem. This becomes prohibitive in time-critical applications, like clinical \ac{CT} imaging, and especially so when the prior model is non-smooth, as in sparsity promoting priors. Another challenge lies in choosing an appropriate prior \cite{Rudin1992,Bredies2010total,Zhang2010bregmanized,Rantala2006wavelet,Bubba2018nonsmooth} and \cite[sec.~3.4]{ActaReview}.}

\rev{Motivated by these shortcomings, recently there have been several efforts in using methods from \rev{deep learning} for reconstruction \cite{ActaReview}. One particular approach trains a deep neural network with a suitable architecture against supervised data using a squared $\ell^2$-loss \cite[sec.~5.1.2]{ActaReview}. This can be seen as an approximate way to compute the average solution (conditional mean estimate). When properly adapted, such data driven approaches considerably outperform purely model based reconstruction techniques regarding \emph{both} reconstruction quality and reconstruction speed \cite{Adler2017}.}


One natural approach is to use \rev{deep learning in the above context} to directly learn the mapping from data to image \cite{Zhu2018automap}. 
Such an approach scales poorly, it requires re-training when data acquisition changes, and it relies on access to huge amounts of training data. Hence, this is not a feasible approach for clinical \ac{CT} where high quality training data is scarce\rev{, as access to projection data is limited}. Another approach is to use \rev{deep learning} as a post-processing tool to improve upon an initial reconstruction. This is computationally feasible as shown in \cite{Kang2017,Jin2017,Ye2018deep}, but such an approach is essentially limited by the information content of the initial reconstruction and the richness of a-priori information learned from training data, which potentially increases bias in the reconstruction.

\emph{Learned iterative reconstruction} methods seek to overcome these drawbacks by combining \rev{deep learning} with a model-based approach. More precisely, the idea is to use a deep neural network architecture for reconstruction that incorporates an explicit handcrafted forward operator and the adjoint of its derivative \cite{Adler2017,Hammernik2018learning,Hauptmann2018TMI,Schlemper2017deep,Li2018nett}. \rev{The idea is to unroll a suitable iterative scheme (usually taken from a model-based approach) that in the limit defines a reconstruction operator \cite[sec.~4.1.9]{ActaReview}}. This yields further improvements to reconstruction quality as compared to \rev{the direct learning or post-processing approaches mentioned before}. Furthermore, including an explicit forward operator improves robustness and generalisability \cite{Boink2019robustness,boink2019TMI}, see also \cite{maier2019learning}. 
Additionally, it also reduces the amount of training data, since networks tend to have \rev{fewer} parameters and the forward operator encodes a major portion of the relations in data that come from the acquisition geometry. 

As already indicated, learned iterative reconstruction methods are typically trained in an end-to-end manner. Hence, the entire unrolled fixed-point scheme is treated as a single network and all its parameters are trained jointly. This provides an optimal set of network parameters under suitable optimisation procedures, but it also comes with two challenges. First, the memory footprint of storing and manipulating the network is too large for most single GPU configurations. Furthermore, during training the loss function is evaluated several times. Each of these involves evaluating the forward operator and its adjoint, or the adjoint of its derivative, which quickly leads to unreasonable training times. Hence, current learned iterative reconstruction algorithms do not scale well to large-scale and higher dimensions, such as fully 3D \ac{CT}. 


One possible solution to address \rev{these computational} challenges is to adopt a greedy approach for training. Here each unrolled iteration in the network is trained separately \cite{Hauptmann2018TMI}. In this way, training of each unrolled iterate and evaluation of the forward operator can be separated, thus rendering a training procedure feasible. On the other hand, such an approach clearly does not represent an optimal selection of network parameters as compared to jointly optimising over all network parameters for all unrolled iterates, as discussed in \cite[sec.~III-A]{Hauptmann2018TMI}. Therefore, such a greedy approach renders a trained network for reconstruction that may fall short in reconstruction quality compared to end-to-end schemes, we refer to \cite{bengio2007greedy} for further discussion on greedy schemes. Additionally, reconstruction times are still comparably slow due to multiple applications of the forward operator. In some cases however, the issue of computation times can be tackled by using faster approximate models \cite{Hauptmann2018MLMIR}, if available, but memory footprint remains an issue.

\rev{In summary, the computational challenges of utilising learned iterative reconstructions are twofold: (i) Managing memory footprint; (ii) Feasible computation and training times. 
As some of these issues could be simply solved with enough computing power, we deliberately consider the case of limited computational resources in this study, instead of utilising large computing facilities, which may not be accessible to a wide range of researchers. Thus, we will limit ourselves here to a single GPU configuration, that 
necessitates the development of 
more memory efficient algorithms. Additionally, to address the second issue we aim to improve reconstruction speed without compromising reconstruction quality.}

\rev{To achieve this we propose a new approach for training learned iterative reconstruction methods} that scales to demanding large-scale tomographic imaging problems. It is a multi-scale scheme that is motivated by the fact that the continuum forward operator can be discretised on various scales. In fact, the ray transform is known to be scale invariant \cite{Natterer2001mathematics}, which defines the forward operator in \ac{CT}, and this consistency across scales can be utilised for reconstruction \cite{Lassas2009,Purisha2017controlled}.
In particular, in our case each unrolled iterate in the network involves discretising the ray transform on a voxalised grid
and the discretisation becomes increasingly fine as the unrolled iterates progress until the final resolution is achieved. Hence, the full high-resolution forward operator is only needed for the final unrolled iterate. Clearly, the approach is not limited to \ac{CT} and readily applies to other tomographic modalities that involve the ray transform. Furthermore, it can be extended to any modality that arises as discretisation from a continuum model, such as MRI or even seismic imaging, in contrast to purely discrete problems.


This paper is structured as follows. In \Cref{sec:learnedIterative} we review common approaches for 
learned reconstructions and discuss possible limitations for 
large-scale applications. In \Cref{sec:multiScale} we introduce 
the notion of multi-scale schemes. In \Cref{sec:3Drecons} we extend the multi-scale scheme to a hybrid network and apply the proposed network to reconstruct from real 
\ac{CBCT} measurements of an organic phantom in 3D. In the following \Cref{sec:compExp} we 
discuss scalability and evaluate performance in comparison to other learned reconstruction methods in 2D for phantoms from human abdominal \ac{CT} scans. In \Cref{sec:Discussion} we discuss extensions and limitations of the proposed multi-scale approaches. Some final conclusions are presented in \Cref{sec:Conclusions}.

\section{Learned reconstructions for tomographic imaging}\label{sec:learnedIterative}
In computed tomography we aim to reconstruct an image of the inside of a patient or object of interest from X-ray measurements. 
Mathematically, this reconstruction task is an inverse problem where we seek to recover the unknown absorption coefficient $\signaltrue \in \RecSpace$ (image) from measured photons $\data\in \DataSpace$ at the sensor (projection data or sinogram) where
\begin{equation}\label{eq:ContinuumModel}
 \data=\ForwardOp(\signaltrue)+\noise.
\end{equation}
Here, $\ForwardOp \colon \RecSpace \to \DataSpace$ is the forward operator, that is assumed to be known, and models how data is generated in absence of noise; $\noise \in \DataSpace$ denotes noise in the observation.

In the following we will assume that $\ForwardOp$ is a linear operator whose sampling is given by the data acquisition geometry, such as the fan beam transform in 2D and cone beam in 3D. 

Reconstruction is typically an ill-posed task, so one needs to use noise-robust inversion procedures. Either by direct reconstruction algorithms, such as filtered backprojection (FBP), or by iterative algorithms that solve a variational problem 
\begin{equation}\label{eqn:variationalModel}
  \signalest := \argmin_{\signal \geq 0} \bigl\{ \mathcal{D}(\signal;\data) + \alpha \mathcal{R}(\signal) \bigr\}.
\end{equation}
\rev{Here, $\signal \mapsto \mathcal{D}(\signal;\data)$ measures the goodness of fit against data $\data$,   $\signal \mapsto \mathcal{R}(\signal)$ is a regularisation term that ensures stability, and $\alpha>0$ is a weighting parameter that regulates the need for stability against the need to fit data.}
These methods tend to perform well, but are ultimately limited by the expressiveness of the hand-crafted regularisation term $\mathcal{R} \colon \RecSpace \to \Real$. Recently, several \rev{researchers} have proposed to either combine direct reconstructions with a learning based post-processing or to learn an iterative algorithm. In the following we give a short overview of possible approaches \rev{that involve the model in the reconstruction process. Either once in \Cref{sec:reconAndPost} and hence rely more on the expressiveness of the learned network, or multiple times in \Cref{sec:learnedIterRecon}, which consequently increases the influence of the model in the reconstruction task.}



\subsection{Reconstruction and post-processing}\label{sec:reconAndPost}
A straightforward approach to use data driven methods in reconstruction is by post-processing an initial reconstruction. More precisely, let $\ForwardOpPseudoInv \colon \DataSpace \to \RecSpace$ be an analytically known reconstruction operator that is proven to be robust. One can then train a convolutional neural network to remove reconstruction artefacts that arise from using $\ForwardOpPseudoInv$ \cite{Kang2017,Jin2017,Schwab2018deep}. 
These artefacts can be quite notable when data is highly noisy or under-sampled. The learned \emph{inverse mapping} is then given as   
\[
 \ForwardOpInvLearned := \Network_\theta \circ \ForwardOpPseudoInv.
\]
The advantage in this approach lies in the analytical knowledge of the reconstruction operator, and hence networks can be designed to exploit structure in reconstruction artefacts. For instance in spatio-temporal problems, if under-sampling artefacts are known to be incoherent in time, the network only needs to learn to combine the spatial information by a temporal interpolation \cite{Hauptmann2019MRM,kofler2019spatio}. On the other hand, for lower dimensional problems, the capacity of the network is essentially limited by the richness of the training data \cite{Hamilton2018deep,Hamilton2019beltrami}. Clearly such an approach is computationally fast since it only requires a single operator evaluation. On the downside, large capacity networks tend to over-fit to the training data and especially so when the training data is scarce. Furthermore, as shown in \cite{Adler2017,Hammernik2018learning,Hauptmann2018TMI,Adler2018}
the results are clearly outperformed by learned iterative reconstruction algorithms that we next describe. 

\subsection{Learned iterative reconstructions}\label{sec:learnedIterRecon}
In learned iterative reconstruction schemes, neural networks are interlaced with evaluations of the forward operator $\ForwardOp$, its adjoint $\ForwardOpAdj$, and possibly other hand-crafted operators. For example, a simple learned gradient-like scheme \cite{Adler2017,Putzky2017recurrent} would be given by
\begin{equation}\label{eqn:classicLG}
	\signal_{i+1} = \Network_{\param_i}\bigl(\signal_i, \ForwardOpAdj(\ForwardOp(\signal_i) - \data)\bigr),\  i=0,\ldots, N-1.
\end{equation}
This defines a reconstruction operator when stopped after $N$ iterates:
\[\ForwardOpInvLearned(\data) := \signal_N
   \quad\text{where $\param = (\param_0,\ldots,\param_{N-1})$} 
\]
and initialisation $\signal_0=\ForwardOpPseudoInv(g)$.
Note that $\Network_{\param_i}$ is a \emph{learned updating operator} for the $i$:th iterate. 
The terminology `gradient-like' comes from the following observation: if we consider minimising $\mathcal{D}(\signal;\data) = \frac{1}{2} \bigl\|\ForwardOp(\signal) - \data \bigr\|_2^2$, then $\Network_\param(\signal, h) := \signal - \theta h$ corresponds to a learned update in a gradient descent scheme, where the step length $\theta$ is the only learned parameter.

The parameters $\param$ in the reconstruction operator $\ForwardOpInvLearned$ are learned by end-to-end supervised training. More precisely, assume one has access to supervised training data $(\signal^{(j)},\data^{(j)}) \in \RecSpace \times \DataSpace$ where $\data^{(j)} \approx \ForwardOp(\signal^{(j)})$. Then an optimal parameter is found by 
\[ \min_{\param} \frac{1}{m} \sum_{j=1}^m     
     \Loss_{\param}(\signal^{(j)},\data^{(j)}) 
\]
where the loss function is given as
\[
    	\Loss_{\param}(\signal,\data) :=\bigl\| \ForwardOpInvLearned(\data)  - \signal \bigr\|_\RecSpace^2
    	\quad\text{for $(\signal,\data) \in \RecSpace \times \DataSpace$}.
\] 
Note here that computing the gradient of the loss function w.r.t. $\theta$ requires performing back-propagation through all of the unrolled iterates $i=0,\ldots, N-1$.

    

In gradient boosting, that follow the greedy training \cite{Hauptmann2018TMI}, the loss function is changed. Instead of looking for a reconstruction operator that is optimal end-to-end, we only require iterate-wise optimality. For the learned gradient scheme above, this amounts to the following loss function for the $i$:th unrolled iterate:
\[
    	\Loss_{\param_i}(\signal_i,\data) =\Bigl\| \Network_{\param_i}\bigl(\signal_i, \ForwardOpAdj(\ForwardOp(\signal_i) - \data)\bigr) - \signal \Bigr\|_\RecSpace^2
\]
where $\signal_i := \Network_{\param_{i-1}}\bigl(\signal_{i-1}, \ForwardOpAdj(\ForwardOp(\signal_{i-1}) - \data)\bigr)$ and initialisation $\signal_0=\ForwardOpPseudoInv(\data)$. These schemes can be viewed as a greedy approach and consequently constitute an upper bound to end-to-end networks. Thus, in the following we seek for a possibility to utilise end-to-end networks for large-scale problems.
    
\section{Multi-scale learned iterative reconstructions}\label{sec:multiScale}
The major limitations when employing learned iterative reconstruction methods for large problems are their prohibitive training times and memory requirements. This is mainly due to the fact that all iterations are performed at full resolution and hence require to evaluate the full scale forward operator for each iterate. To overcome this limitation we propose a multi-scale scheme.

\begin{figure*}[th!]
\centering
\footnotesize
\begin{picture}(500,100)

\put(0,0){\includegraphics[width=500pt]{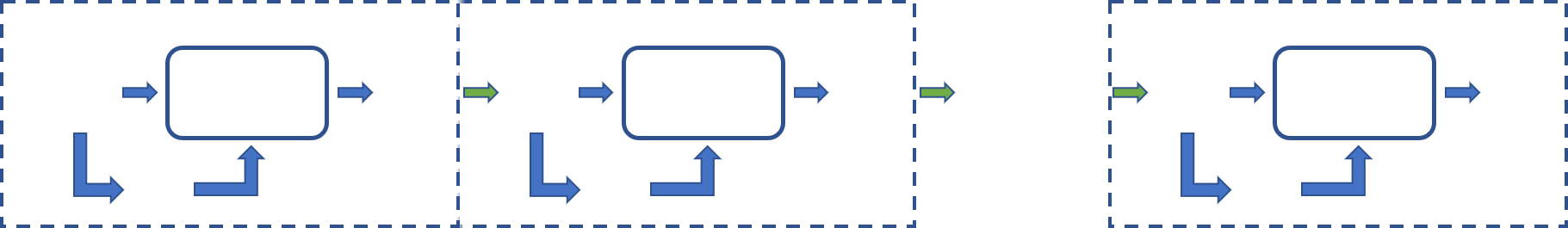}}

\put(55,80){Iterate 0} 
\put(35,62.5){Discretisation Space $S_0$} 
\put(20,42){$\widetilde{\signal}_0$}
\put(40,9){$\nabla\mathcal{D}_0$}
\put(73,42){$\Network_{\theta_0}$} 
\put(125,42){${\signal}_0$}

\put(205,80){Iterate 1} 
\put(185,62.5){Discretisation Space $S_1$} 
\put(148,47){$\tau_1$}
\put(166,42){$\widetilde{\signal}_1$}
\put(186,9){$\nabla\mathcal{D}_1$}
\put(220,42){$\Network_{\theta_1}$} 
\put(270,42){${\signal}_1$}

\put(293,47){$\tau_2$}
\put(320,40){{\large{$\ldots$} }}

\put(410,80){Iterate N} 
\put(390,62.5){Discretisation Space $S_N$} 
\put(355,47){$\tau_N$}
\put(370,42){$\widetilde{\signal}_{N}$}
\put(393,9){$\nabla\mathcal{D}_{N}$}
\put(425,42){$\Network_{\theta_{N}}$} 
\put(473,42){$\signal_N$}

\end{picture}
\caption{\label{fig:MSLGD_graph} Visualisation of the \ac{MS-LGS} as outlined in 
\cref{alg:MSLGD}. Each iteration is performed on their respective discretisation space, where the gradient $\nabla\mathcal{D}_{i}:=\nabla \mathcal{D}_i(\signal_i;g)$ is computed and the update is performed by the network $\Network_{\theta_{i}}$. After each update an up-sampling by $\tau_i$ to the next finer space is performed until the final resolution $S_N$ is achieved.}
\end{figure*}
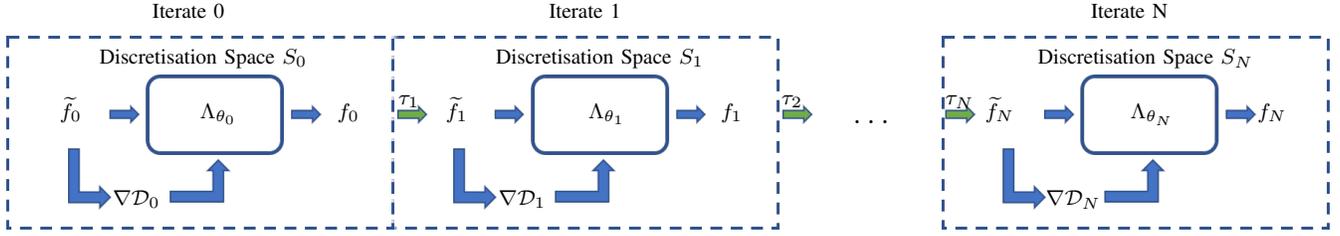

\subsection{Discretisation sequence}
In the inverse problem in \cref{eq:ContinuumModel}, both the unknown image $\signaltrue$ and data $\data$ are considered as continuum objects, which in imaging are typically represented by real-valued functions defined on some domains. 
In reality discrete data is recorded through a measurement device and we can only compute a digitised version of the unknown $\signaltrue$. 
By discretisation we refer loosely to the procedure for defining a finite dimensional version of \cref{eq:ContinuumModel} that is given by the finite sampling of the data and the digitisation of $\signaltrue$.
Likewise, a \emph{discretisation sequence} is a finite sequence of discretisations that start from a coarse discretisation and is successively refined towards the desired finest resolution. 
The refinement and coarsening of the discretisation is through specific up- and down-sampling schemes that will be defined later.
Consequently, motivated by the discretisation invariance of the ray transform, we aim to iteratively increase the resolution of our reconstructions. 
For that purpose, let $S_0,\dots,S_N$ denote a fixed sequence of discretisations of $\RecSpace$ and $\DataSpace$ that increase in resolution through subsequent up-sampling. In the following we will associate each iterate $f_i$ with such a discretisation space $S_i$. 
Stated more formally, a discretisation sequence is given by 
\[ S_i:=\RecSpace_{i} \times \DataSpace_{i} 
   \quad\text{for $i=0,\ldots,N$.} 
\]
Here, $\RecSpace_{i} \subset \RecSpace$ is a finite dimensional subspace with \rev{dimension $\dim(\RecSpace_i) \leq \dim(\RecSpace_{i+1})$}. Likewise, $\DataSpace_{i} \subset \DataSpace$ with \rev{$\dim(\DataSpace_i) \leq \dim (\DataSpace_{i+1})$}. Furthermore, let $\{\signal_{i},\data_i\} \in S_i$ denote the reconstructed image and data in each discretisation space. 
In the following we will need a projection operator in the data space $\ProjectScale_i \colon \DataSpace \to \DataSpace_i$, for $i=0,\dots,N$, and an up-sampling operator in the image space $\UpSample_i \colon \RecSpace_{i-1} \to \RecSpace_{i}$, for $i=1,\dots,N$. Whereas the projection operator maps the data into the respective discretisation space, the up-sampling operator maps the reconstruction in the $i$:th discretisation space to the subsequent one in the discretisation sequence. Note that if \rev{$\dim(X_{i-1})=\dim(X_i)$}, then the up-sampling reduces to the identity $\tau_i=\bf{id}$.


The discretisation sequence $S_0, \ldots, S_N$ defines as well a sequence of discretised versions of the inverse problem in \cref{eq:ContinuumModel}. More precisely, for each discretisation $S_i$ we obtain the corresponding inverse problem of recovering $\signaltrue_i\in \RecSpace_{i}$ from finitely sampled data $\data_i \in \DataSpace_{i}$ where
\[
  \data_i = \ForwardOp_i(\signaltrue_i) + \noise_i
\]
with $\noise_i$ denoting the noise in data and $\ForwardOp_i \colon \RecSpace_{i} \to \DataSpace_{i}$ denoting the corresponding forward operator. Similarly, we have $\ForwardOp^*_i \colon \DataSpace_i\to \RecSpace_i$ for the adjoint and $\ForwardOp_i^\dagger \colon \DataSpace_i\to \RecSpace_i$ for the pseudo-inverse on the discretisation space $S_i$, e.g. the filtered backprojection in 2D or \ac{FDK} in 3D. 
With these concepts we can now formulate the multi-scale iterative reconstructions schemes.


\subsection{A multi-scale learned gradient scheme}
The underlying principle of the proposed multi-scale scheme is to start at the coarsest discretisation space $S_0$ and after each iterate we up-sample until we obtain the reconstruction in the final discretisation space in the desired full-resolution. This way each iterate has its own discretisation space and hence the number of iterations we perform is $N+1$, equal to the number of discretisation spaces.
Since we aim to train the algorithm end-to-end, this maximum number of iterations has to be fixed. For each iterate we then compute the gradient in the corresponding discretisation space $\nabla \mathcal{D}_i(\signal_i;g)\in S_i$ given by
\begin{equation}\label{eqn:scaleGrad}
\nabla \mathcal{D}_i(\signal_i;g):= \ForwardOp^*_i\bigl(\ForwardOp_i(\signal_i) - \pi_i(\data)\bigr).
\end{equation}
Following the structure of learned gradient schemes \cref{eqn:classicLG}, we perform a learned update with the current reconstruction $\signal_i$ and the corresponding gradient $\mathcal{D}_i(\signal_i;g)$, followed by an up-sampling to the next finer resolution,
\[
\begin{cases}
\signal_{i}=\Network_{\theta_i}\bigl(\widetilde{\signal}_i,\nabla \mathcal{D}_i(\widetilde{\signal}_i;\data) \bigr)
&\\[0.5em]
\widetilde{\signal}_{i+1}=\UpSample_{i+1}(\signal_i). 
\end{cases}
\]
The full \ac{MS-LGS} is summarised in \cref{alg:MSLGD} and a schematic is illustrated in \cref{fig:MSLGD_graph}.



\begin{algorithm}
	\caption{\Acf{MS-LGS}}
	\label{alg:MSLGD}
	\begin{algorithmic}[1]
        \For{$i=0,\dots,N$}
            \If{$i = 0$}
		        \Let{$\widetilde{\signal}_0$}{$\ForwardOp_0^\dagger \pi_{0}(\data)$}
            \Else
                \Let{$\widetilde{\signal}_i$}{$\tau_i(\signal_{i-1})$}
            \EndIf
            \Let{$\signal_{i}$}{$\Network_{\theta_i}\left(\widetilde{\signal}_i,\nabla \mathcal{D}_i(\widetilde{\signal}_i;\data) \right )$}
        \EndFor
        \Let{$\signal^*$}{$\signal_N$}
	\end{algorithmic}
\end{algorithm}

\subsubsection{Including a filtered gradient}\label{sec:MSLFGS}
Let us first note, that the up-sampling operator in each iteration restricts the high frequency components that can be present after up-sampling. Additionally, the normal operator $\ForwardOp^*\ForwardOp$ is known to be smoothing of order 1 \cite{Natterer2001mathematics}, which means, effectively, that 
any high frequency components in the final reconstruction can only be introduced by the network, similarly to the role of the regulariser in classical variational techniques. Thus, to complement the information for the network, we consider a version of \ac{MS-LGS} with an additional filtered gradient that retains higher frequencies. That means we do not only compute the classic gradient $\nabla\mathcal{D}_i(\signal_i;g)$ in each iteration, but additionally a filtered version by substituting the adjoint with the filtered backprojection, or \ac{FDK} in 3D, 
\begin{equation}\label{eqn:filteredGrad}
\nabla^\dagger \mathcal{D}_i(\signal_i;\data):= \ForwardOp^\dagger_i\bigl(\ForwardOp_i(\signal_i) - \pi_i(\data)\bigr).
\end{equation}
A similar approach has been studied earlier for classic iterative methods in \cite{Gao2016FDK}. In our case the filtered gradient will be computed additionally to the classic gradient \cref{eqn:scaleGrad} and hence this will increase the computational cost by the application of one filtered backprojection in each step, but, as can be seen later, 
improves reconstruction quality. For notational convenience, we will denote the set of inputs to the network in each scale by
\begin{equation}\label{eqn:signalSet}
[\widetilde{\signal}_i] := \left\{\widetilde{\signal}_i,\nabla  \mathcal{D}_i(\widetilde{\signal}_i;\data),\nabla^\dagger	\mathcal{D}_i(\widetilde{\signal}_i;\data)\right\}.
\end{equation}
In the resulting scheme, \ac{MS-LFGS}, with the additional computation of the filtered gradient we then have the update equations
\[
\signal_{i} \leftarrow   
  \Network_{\theta_i}\bigl( 
    [\widetilde{\signal}_i] 
  \bigr)
\]
instead of line 7 in \cref{alg:MSLGD}.

\subsubsection{Computational cost}\label{sec:CompCost}


Concerning the total computational cost: Due to sub-sampling on the coarser discretisation spaces the computation of projections is essentially governed by the computations on the final resolution. If we assume that the computational cost of evaluating the network $\Network_{\theta_i}$ is negligible in comparison to the forward and adjoint operator (or pseudo-inverse), then the total computational complexity is governed by the cost of the operator at the finest scale. 

Formally, the total computational cost can be roughly estimated as follows. Let us assume that at each scale we double each dimension, then the number of voxels scale by $2^d$.
Thus, the computational cost on each scale increases in the same manner and 
the estimated total computational cost on all scales can be bounded by a geometric series
\begin{equation}\label{eqn:geometricSeries}
C_d:=\sum_{k=0}^{\infty}\left(\frac{1}{2^d}\right)^k = \frac{1}{1-1/2^d}.
\end{equation}
For $d=2$ we have $C_2= 4/3$ and $C_3=8/7$ for $d=3$. \rev{We note, that the same estimate applies to memory requirements of the multi-scale scheme.} This emphasises that the proposed approach is especially suitable for higher dimensional applications, since the computational cost on the course discretisation spaces becomes neglectable, as we will see in the next section for an application to 3D cone beam \ac{CT}.


\section{Reconstruction of 3D cone beam measurements}\label{sec:3Drecons}
Let us now discuss the reconstruction task from three dimensional cone beam measurements. We note that due to the structure of the multi-scale approaches, the reconstruction quality will essentially depend on the expressibility of the last layer and hence it is only reasonable to make the last iterate as informative as possible. To achieve scalability with an expressive network at the last iterate, we propose to combine \ac{MS-LFGS}, as described in \Cref{sec:MSLFGS}, with the established U-Net architecture \cite{Ronneberger2015} with the addition that the gradient information is reused in each scale of U-Net.
This network is specifically designed to utilise the previously computed information across all-scales.

\subsection{Cone beam measurement data}
We evaluate the applicability of the proposed networks to reconstructions in 3D with an application to \ac{CBCT}. For this purpose we utilise a database provided by the FleX-ray lab at \rev{Centrum Wiskunde \& Informatica} \cite{der2019cone}, consisting of 42 walnuts scanned in a custom made $\mu$CT. For each target there are 3 separate scans consisting of 1201 angles \rev{with uniform increment of $0.3^\circ$ and varying source locations at the top, middle, and bottom of the target. That is for a mean target size of 30mm the scanning positions are at -15mm, 0, 15mm with respect to the central slice. As these three scans result in different cone beam artefacts, they are combined to create a reference ground-truth reconstruction of size $501^3$ negating the cone beam artefacts.
We refer to \cite{der2019cone} for further details on the scanning setup and geometry.}

\rev{For our experiments we utilise the central scanning position at the centre of the target, with a source-to-target distance of 66mm and source-to-detector distance 199mm.
We select 60 uniformly spaced angles, resulting in an angular increment of $6^\circ$. Additionally we \rev{down-sample} both, measurements and ground-truth reconstruction, by a factor of 3. This results in a reconstruction size of $168^3$, where each of the 60 projections is of size $256\times 324$. We note that we chose the maximum reconstruction size possible under the memory constraints of this study.}

The supplied data is given as linearised measurements, thus we will use the linear projection model as our forward operator
\begin{equation}\label{eqn:rayTrafo}
\ForwardOp(\signal)(\ell) = \int_\ell \signal(x)\, \text{d}x, 
\quad\text{for $\ell \in \mathcal{M}$,}
\end{equation}
where $\mathcal{M}$ is the \rev{three dimensional manifold of lines in $\Real^3$} defined by the \rev{cone beam measurement geometry described above.}

\begin{figure*}[th!]
\centering
\footnotesize
\begin{picture}(500,250)

\put(0,0){\includegraphics[width=460pt]{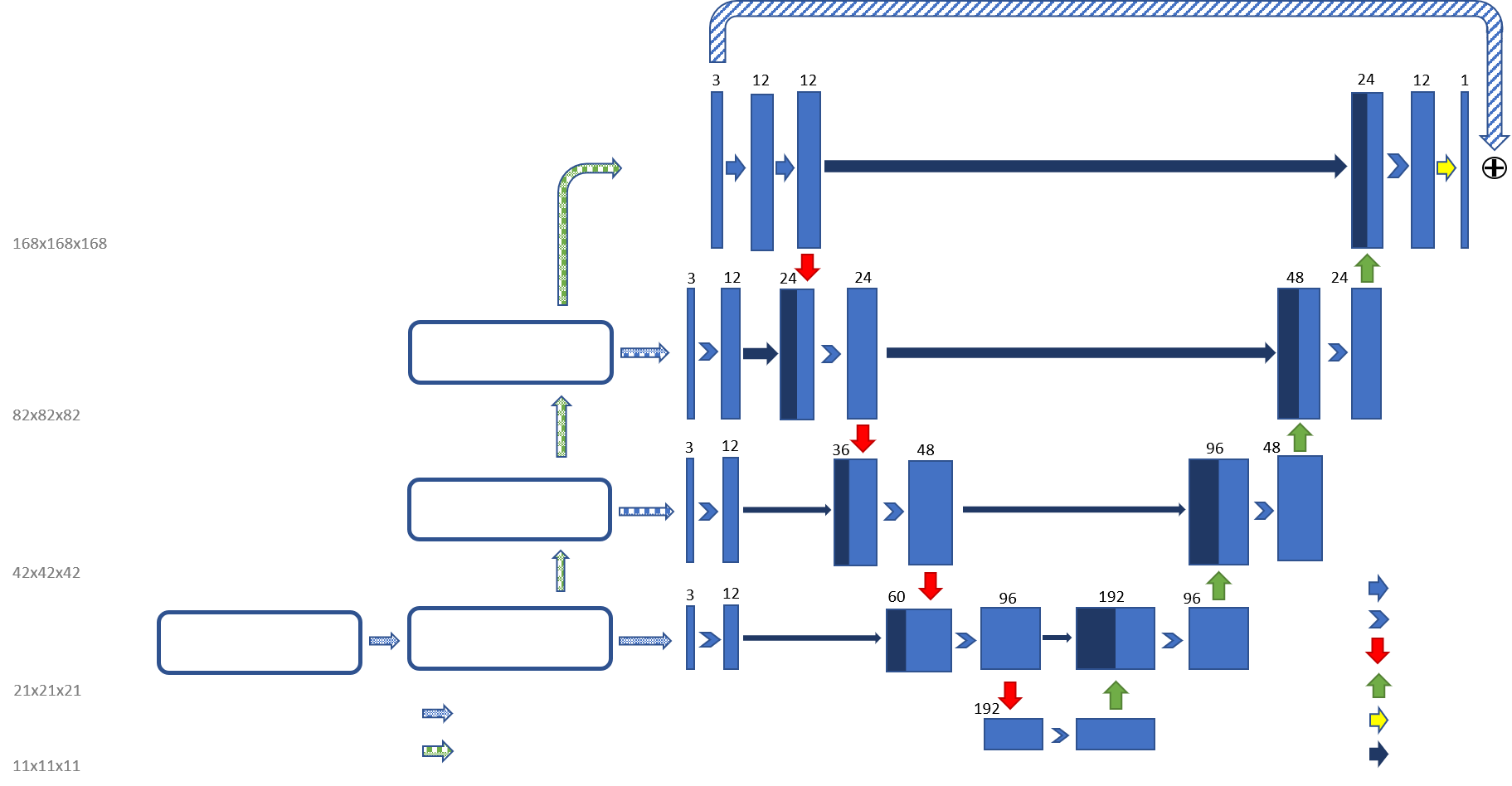}}

\put(290,245){Residual connection: $\widetilde{\signal_4}$}
\put(140,19){\footnotesize Gradient computation}
\put(140,8){\footnotesize Upsample + Gradient computation}

\put(52,42){$\Network_{\theta_0}([\widetilde{\signal}_0])=\signal_0$}  
\put(130,42){$\Network_{\theta_1}([\widetilde{\signal}_1])=\signal_1$} 
\put(130,81){$\Network_{\theta_2}([\widetilde{\signal}_2])=\signal_2$} 
\put(130,130){$\Network_{\theta_3}([\widetilde{\signal}_3])=\signal_3$} 
\put(192,186){\footnotesize $[\widetilde{\signal}_4]=$} 
\put(460,186){\footnotesize $=\signal_4$} 

\put(4,170){\footnotesize Scale 4}
\put(4,117){\footnotesize Scale 3}
\put(4,69){\footnotesize Scale 2}
\put(4,33){\footnotesize Scale 1}
\put(4,12){\footnotesize Scale 0}

\put(430,58){\footnotesize ReLU(conv$_{3\times 3 \times 3}$)}
\put(430,48){\footnotesize $2\times$ReLU(conv$_{3\times 3 \times 3}$)}
\put(430,38){\footnotesize maxpool$_{2\times 2 \times 2}$}
\put(430,28){\footnotesize convt$_{2\times 2 \times 2}$}
\put(430,18){\footnotesize conv$_{1\times 1 \times 1}$}
\put(430,8){\footnotesize concat}
\end{picture}
\caption{\label{fig:dUnet} The proposed $\partial$U-Net architecture for multi-scale learned iterative reconstructions of \ac{CBCT} reconstructions in 3D. The left part of the network consists of a \ac{MS-LFGS}, which uses a U-Net on the right in the final iterate. Additionally, the output and corresponding gradient information of each iterate is re-used in the respective scale of the U-Net. 
}
\end{figure*}
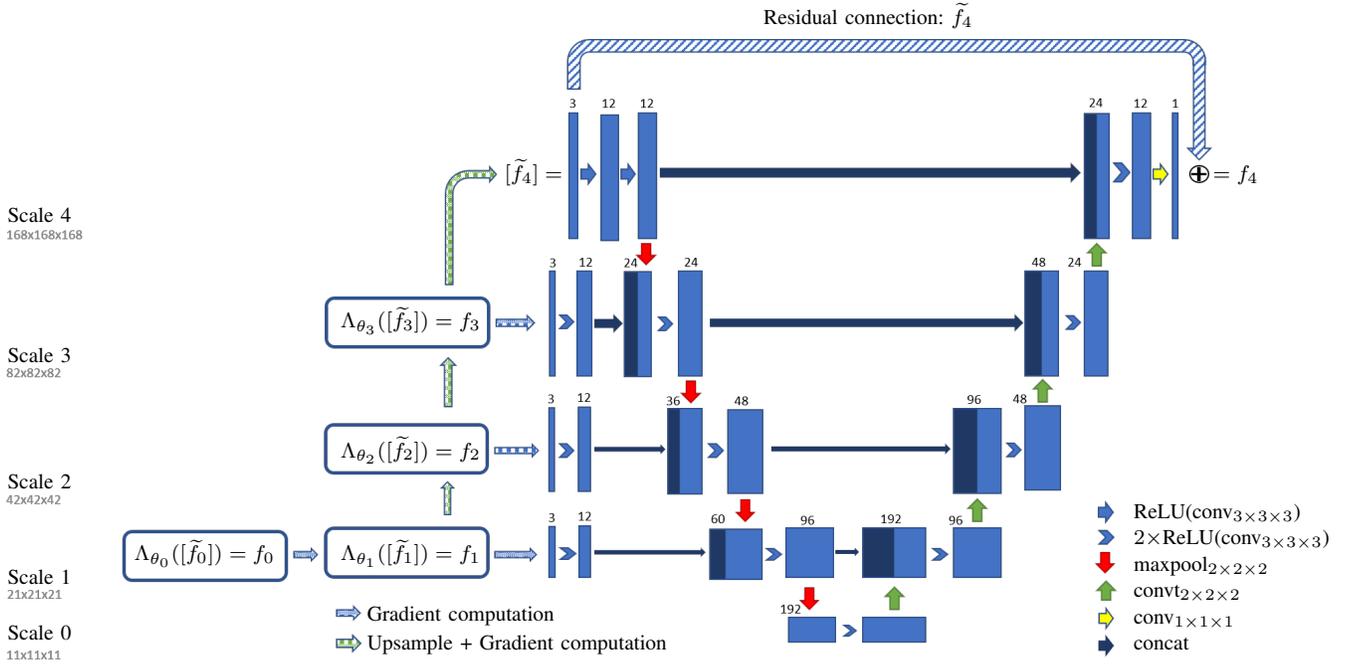

\subsection{A hybrid multi-scale network: $\partial$U-Net}
As the final reconstruction quality in the multi-scale scheme is primarily dependent on the last iterate operating on the final resolution, it is advisable to make this last iterate as 
expressive as possible without significantly increasing bias in the reconstructions. For this purpose we propose an across-scales network, that is essentially a combination of \ac{MS-LFGS} and U-net that utilises the computed gradient information across all scales;
in the following we will call this architecture $\partial$U-Net. Details of the network design are discussed next. 

\subsubsection{Implementation details}
The resulting $\partial$U-Net architecture chosen for the application to \ac{CBCT} is illustrated in \Cref{fig:dUnet}. 
We have chosen the number of iterates as $N+1=5$; for the corresponding discretisation spaces, we fix the resolution of the finest desired reconstruction space as $\RecSpace_{N}=\Real^{n\times n\times n}$\rev{, with $n=168$.} The coarser resolutions are then obtained by reducing the resolution for each downsampling by a factor of 2 in each dimension until scale 1, here scale 0 has the same resolution to avoid overfitting due to very small image sizes in the first iterate.
Thus, the coarsest scale is obtained by 3 times downsampling, that is a factor of 8 per dimension and hence the total image size is reduced by a factor of 512.
In the projection space, we keep the number of angles at 60 for each scale, but downsample the detector size by the same factor as the image size, \textit{i.e.} reducing each dimension by factor 2 until scale 1.



The mapping $\pi_i$ to the coarser scale is implemented by an area mean, the up-sampling with $\tau_i$ is performed by trilinear interpolation. After each network update, we compute the set of filtered and classical gradient as in \cref{eqn:signalSet} for the current scale, that is 
\begin{equation}\label{eq:extraGrad4dUnet}
    [{\signal}_i] = \Bigl[ \Network_{\theta_i}\bigl([\widetilde{\signal}_i] \bigr) \Bigr],
\end{equation}
as well as the gradient set of the up-sampled output $\widetilde{\signal}_{i+1}=\tau_{i+1}(\signal_i)$. Where the former gradient set in \cref{eq:extraGrad4dUnet} is passed to U-Net in the respective scale, subsequently expanded by a double convolutional layer and then concatenated with the result of the max-pooling in U-Net,
    and the latter gradient set of the up-sampled output, i.e.  $\widetilde{\signal}_{i+1}=\tau_{i+1}(\signal_i)$, is used for the next iterate in the gradient scheme. Here the sub-networks are given in a ResNet style following \cite{Adler2017,Adler2018}. Specifically, we chose a double convolutional layer with 12 channels and a final layer with 1 output channel. The output is then given by a residual update 
\begin{equation}\label{eqn:resUpdates}
\Network_{\theta_i}\bigl([\widetilde{\signal}_i] \bigr) = \signal_i + s_i\mathcal{G}_{\widetilde{\theta}_i}\bigl([\widetilde{\signal}_i]\bigr),
\end{equation}
where $\mathcal{G}_{\widetilde{\theta}_i}$ denotes the chosen architecture for the updates, i.e. the the three convolutions, and $s_i$ is a learnable step size initialised by 0 following \cite{zhang2018residual}. The learnable parameters in each iterate are then given by $\theta_i=\{s_i,\widetilde{\theta}_i\}$.

All algorithms, including reference methods, are implemented in Python using PyTorch \cite{Paszke2017automatic} for the networks. The image and projection spaces are implemented with ODL (Operator Discretization Library) \cite{Adler2017ODL} using ASTRA \cite{Van2016Astra} as \rev{back-end for evaluating the ray transform and its adjoint}. Training details and parameter choices will be stated in the following sections.

{\newcommand{\showpic}[2]{%
\begin{tikzpicture}
\draw (0,6) node [anchor=south] {\phantom{f}#1\phantom{g}};%
\draw (0,0) node [anchor=north] {\includegraphics[width=3cm]{walnut/#2_xy}};%
\draw (0,3.1) node [anchor=north] {\includegraphics[width=3cm]{walnut/#2_yz}};%
\draw (0,6.2) node [anchor=north] {\includegraphics[width=3cm]{walnut/#2_xz}};%
\end{tikzpicture}\hspace*{-2mm}%
}

\begin{figure*}[th!]
\centering
\showpic{Phantom}{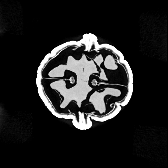}%
\showpic{\acs{FDK}}{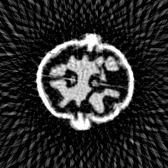}%
\showpic{U-Net denoiser}{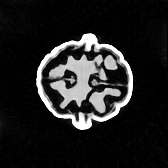}%
\showpic{\acs{MS-LFGS}}{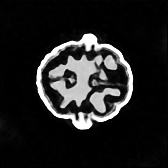}%
\showpic{$\partial$U-Net}{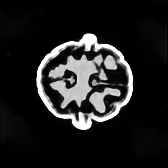}%
\put(-440,25){\rotatebox{90}{xy-plane}}
\put(-440,120){\rotatebox{90}{yz-plane}}
\put(-440,210){\rotatebox{90}{xz-plane}}
\caption{\label{fig:walnutRecs} Reconstructions of the Walnut used for testing from 60 angles \rev{and resolution $168^3$}. Reconstructions are compared to the phantom computed from a total of 1200 angles and three scanning positions to negate cone beam artefacts. The reconstruction by \acs{FDK} is computed with Hann filter and frequency scaling $h=0.6$ (PSNR=26.95). The proposed algorithm $\partial$U-Net (PSNR=34.69) is compared to post-processing by U-Net (PSNR=34.62) and the multi-scale approach \ac{MS-LFGS} (PSNR=34.13) using a mini U-Net in each scale. }
\end{figure*}}

\begin{table*}[ht!] 
\small
  \caption{Quantitative measures \rev{and computational resources} for \ac{CBCT} reconstructions of the walnut data from 60 angles. Computed values are given for the test data in comparison to the ground-truth from full measurements. \rev{Additionally, we present estimated benchmark values for \acs{LGS}.} }
  \begin{center}
    \begin{tabular}{l|r|r|r|r|r|r}
&{\sc PSNR} &{\sc SSIM} & {\sc Train} &{\sc Exec.} &{\sc Parameter} &{\sc Memory}    \\
    \midrule
  {\sc \acs{FDK} } &    26.95 &   0.424  &  $\sim$1m & 260ms & 1 & 723MB \\ 
  \rev{{\sc U-Net denoiser}}&    34.62 &   0.910  & 4h29m & 528ms & 6.3$\cdot10^6$ & 6097MB \\
  {\sc \acs{MS-LFGS} (ResNet)}&   32.98 &   0.878  &  5h04m & 526ms & 2.4$\cdot10^4$ & 2547MB \\
  {\sc \acs{MS-LFGS} (mini U-Net)}&  34.13 &  0.903  &  7h02m &   645ms & 2.1$\cdot10^5$ & 4853MB  \\
  {\sc $\partial$U-Net}& \bf 34.69 & \bf 0.914 & 7h28m & 795ms & 4.1$\cdot10^6$ & 5313MB   \\
  \rev{{\sc \acs{LGS} (5 iter., estimated)}}& \bf -- & \bf -- & $\sim$25h & 2.5s & $\sim 2\cdot10^5$ & $\sim$ 12500MB    \\
    \bottomrule
    \end{tabular}%
  \label{table:walnutQuant}%
  \end{center}
\end{table*}%

\subsection{Reconstructions}
Additionally to reconstructions with the proposed $\partial$U-net, we will compare the quality to reconstruction with \ac{FDK}
followed by post-processing with U-Net, following \cite{Jin2017}, as well as a reconstruction with the basic \ac{MS-LFGS} as described in \Cref{sec:multiScale}. We note that this is essentially an ablation study on how each part performs separately. The U-Net architecture follows the same scheme as outlined in \Cref{fig:dUnet}, with the difference that the initial channel width is 16 and doubled in each scale, leading to slightly more parameters. For \ac{MS-LFGS} we chose two variants here, one that is based as well on a ResNet architecture as used in the $\partial$U-Net and a second variant, where all sub-networks $\mathcal{G}_{\widetilde{\theta}_i}$ in \eqref{eqn:resUpdates} are given by a down-scaled version of U-Net, which we call mini U-Net, similarly to what has been used in \cite{Hauptmann2018MLMIR}. This mini U-Net consists of only 2 scales (one max-pool layer), instead of the classic 4, and an initial channel depth of 12 on the first
scale to be conforming with the $\partial$U-Net. All updates in the iterate schemes are performed following the residual updates in \cref{eqn:resUpdates}.

To make the comparison uniform for all test cases we performed training for all algorithms in the same manner. In particular we chose Adam as the optimiser with an $\ell^2$-loss to the ground-truth; 
each network is trained for 10,000 iterations with one training sample per minimisation step. The initial learning rate is set to $10^{-3}$ with a cosine decay. These choices have shown to perform well for all presented algorithms. 

For training we have chosen 40 out of the 42 walnuts, which leaves 2 for validation and testing. The obtained reconstructions for the test walnut (number 41) are shown in \Cref{fig:walnutRecs}. It can be seen, that all learned methods are capable of successfully suppressing the cone beam artefacts in comparison to the \ac{FDK} reconstruction.

\subsection{Quantitative results}
Visually all three learned reconstructions perform well and produce an informative reconstruction from just 60 projection angles. To compare the reconstructions in more detail, we have computed quantitative measures shown in \Cref{table:walnutQuant}, specifically PSNR and SSIM with respect to the provided ground-truth image. Additionally, we provide training and execution times for all algorithms, number of parameters and needed memory for evaluation of the trained network.

The results suggest that the basic multi-scale approach is not competitive in terms of PSNR and SSIM. As we have indicated earlier, this is most likely due to the limited expressiveness of the final network. This can be clearly seen by the comparison of \ac{MS-LFGS} based on ResNet and the mini U-Net for each iterate, as increasing the depth of the networks improves reconstruction quality clearly. 
In particular, the proposed $\partial$U-Net, that combines the \ac{MS-LFGS} architecture with a U-Net in the final iterate, improves reconstruction quality further and slightly outperforms the established post-processing and denoising by U-Net approach. 

Concerning training and execution times, clearly U-net is fastest to train and execute, roughly taking double the time of \ac{FDK}.
It is noteworthy that the iterative approaches only add a slight overhead in execution time, where \ac{MS-LFGS} using a ResNet structure is even faster. The most computationally expensive algorithm is $\partial$U-Net\rev{, but has only} an overhead of 50\% to the basic U-Net. This emphasises the excellent scalability of the multi-scale approaches in 3D.

\rev{In comparison, the basic \ac{LGS} as described in \Cref{sec:learnedIterRecon} with 5 iterates and a ResNet structure would require roughly 5 times the resources, in terms of memory and computation times, see \Cref{table:walnutQuant} for the estimated values. Clearly, one would not only need more computing power to train the algorithm, but also reconstruction times fall short to the multi-scale approaches}

\subsection{\rev{Robustness}}

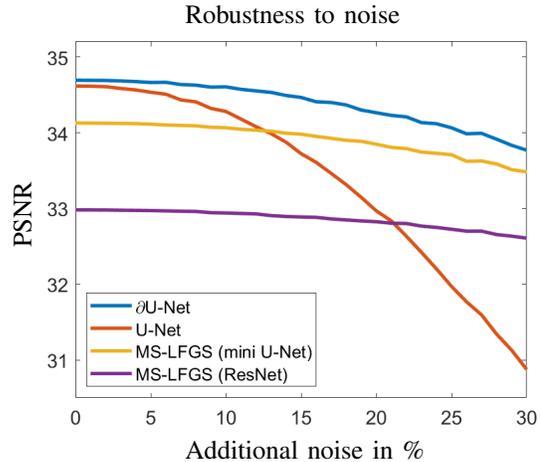
\begin{figure}[t!]
\centering
\begin{picture}(200,170)

\put(-10,10){\includegraphics[width=220pt]{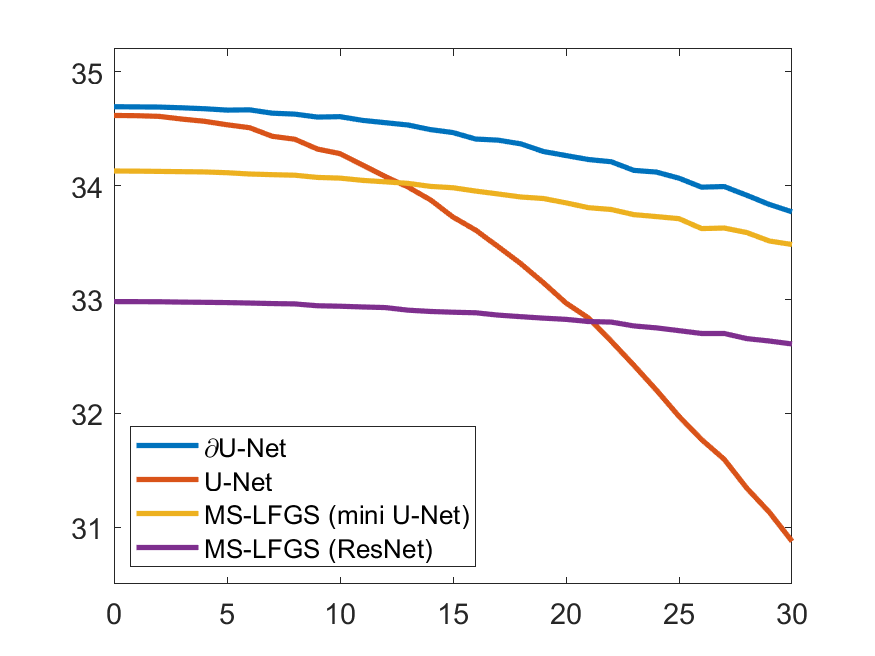}}
\put(60,170){Robustness to noise} 
\put(-5,85){\rotatebox{90}{PSNR}}
\put(60,5){Additional noise in \%}
\end{picture}
\caption{\label{fig:noiseRobust} Robustness study with respect to additional noise in the test data. Specifically, normally distributed random noise is added to the projection data and reconstruction quality is evaluated for all algorithms under consideration.}
\end{figure}

Even though the reconstruction quality of $\partial$U-Net might only slightly outperform the denoising with U-Net, it provides a scalable model-based iterative reconstruction technique. This is of particular importance for applications where training data is scarce and objects might vary, as model-based iterative reconstructions have been shown to be more robust with respect to perturbations in the data and geometry, as demonstrated in several studies \cite{Hauptmann2018TMI,Boink2019robustness,boink2019TMI}\rev{, see also \cite{maier2019learning} for a theoretical discussion.} 
To emphasise this point, we have performed a robustness study with respect to noise. As the training data was given by real projection data it contained a natural noise component. For the robustness study, we have added additional normally distributed noise to the projection data for the test set and recorded the PSNR values of the reconstructions. The results of this experiment are illustrated in \Cref{fig:noiseRobust}. It can be seen that all model-based iterative approaches are more robust with respect to additional noise, whereas post-processing with U-Net does deteriorate much quicker. It is also interesting to note, that $\partial$U-net does show similar robustness as the \ac{MS-LFGS} approaches, but under higher noise starts to deteriorate also a bit faster, which can be expected as it is a hybrid network combining both approaches.

\section{Comparative study in 2D}\label{sec:compExp}

In this section we aim to evaluate the performance of the proposed $\partial$U-Net and multi-scale schemes in comparison to learned gradient schemes as in \cite{Adler2017} that operate in each iterate on the full resolution. As these approaches do not scale well to 3D we restrict ourselves here to two dimensions. We will first examine scalability on simulated data and then evaluate reconstruction performance with realistically generated data from human phantoms supplied for the 2016 AAPM Low Dose CT Grand Challenge.

\subsection{Implementation}
Let us first discuss the implementation choices for the multi-scale schemes. As in the previous section, we fix the number of iterations to $N+1=5$.
To create the discretisation spaces, we fix the resolution of the finest desired reconstruction space as $\RecSpace_{N}=\Real^{n\times n}$. The coarser resolutions are then obtained by reducing the resolution for each downsampling by a factor of 2 in each dimension. That means, the coarsest scale is obtained by 4 times downsampling which reduces the data size in 2D by a factor of 256. In this part, we reduce the amount of angles by a factor of 2 as well, the projection resolution is determined for each scale separately to fully cover the domain.
Following the study in 3D, the mapping $\pi_i$ to the coarser scale is implemented by an area mean, whereas the up-sampling with $\tau_i$ is performed here by bilinear interpolation.

We will restrict the network architectures in this section to learned gradient schemes with a mini U-net as the sub-network. As this choice has shown to be more competitive for the reconstruction of the walnut data in 3D. For the $\partial$U-net, we follow the architecture outlined in \Cref{fig:dUnet}, where we adjust the channel width to 16 in the first scale of the U-Net, this also applies to the sub-networks used in the iterative multi-scale part.


{\newcommand{\showpic}[2]{%
\begin{tikzpicture}[spy using outlines={circle, magnification=4, 
size=2cm, connect spies}]%
\draw (0,0) node [anchor=south, yshift=-5pt] {\phantom{f}#1\phantom{g}};%
\draw (0,0) node [anchor=north] {\includegraphics[width=2.75cm]{recons/#2}};%
\spy on (0,-1.15cm) in node at (-.2cm, -3.5cm);
\end{tikzpicture}\hspace*{-2mm}%
}

\begin{figure}[t!]
\centering
\showpic{Phantom}{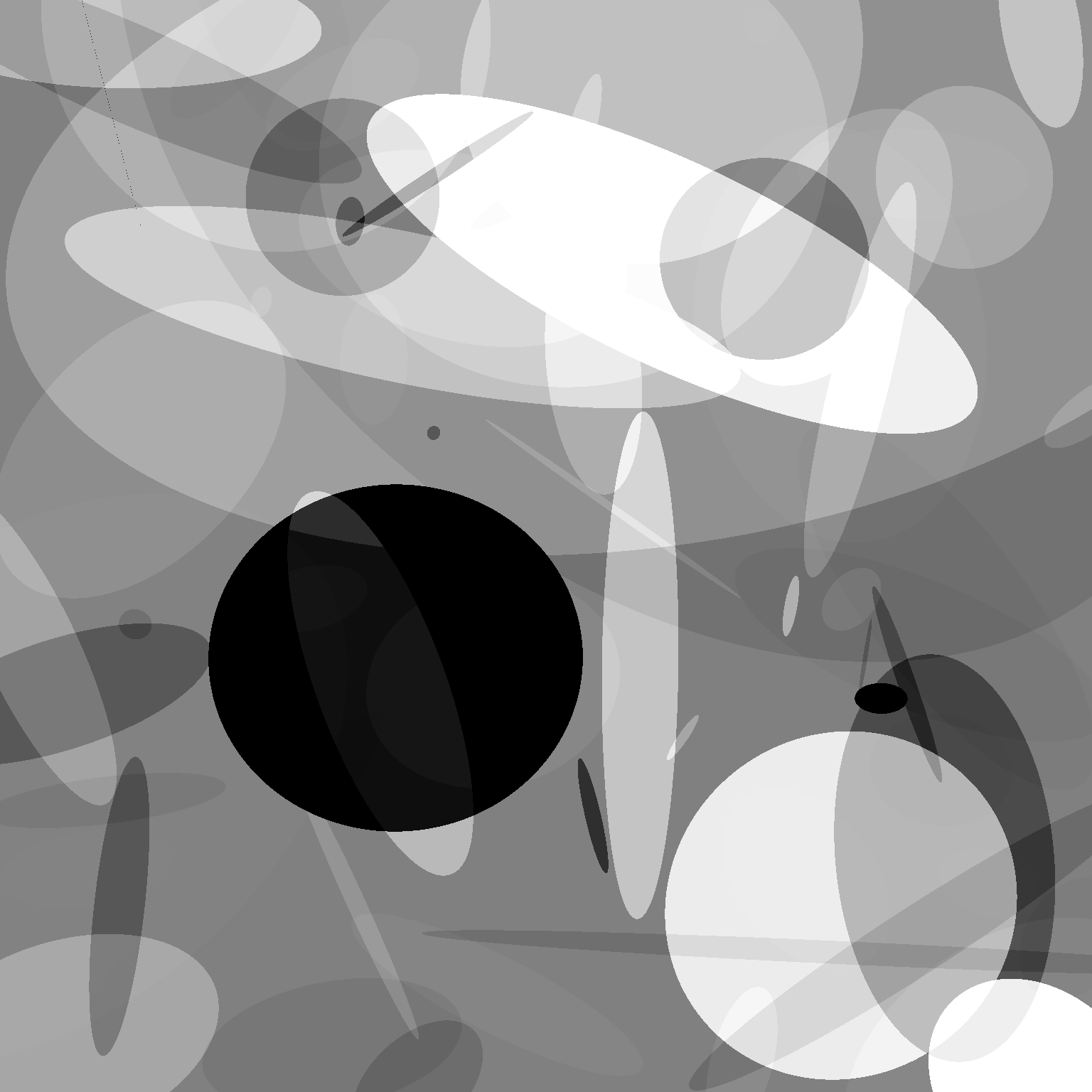}%
\hfill
\showpic{FBP}{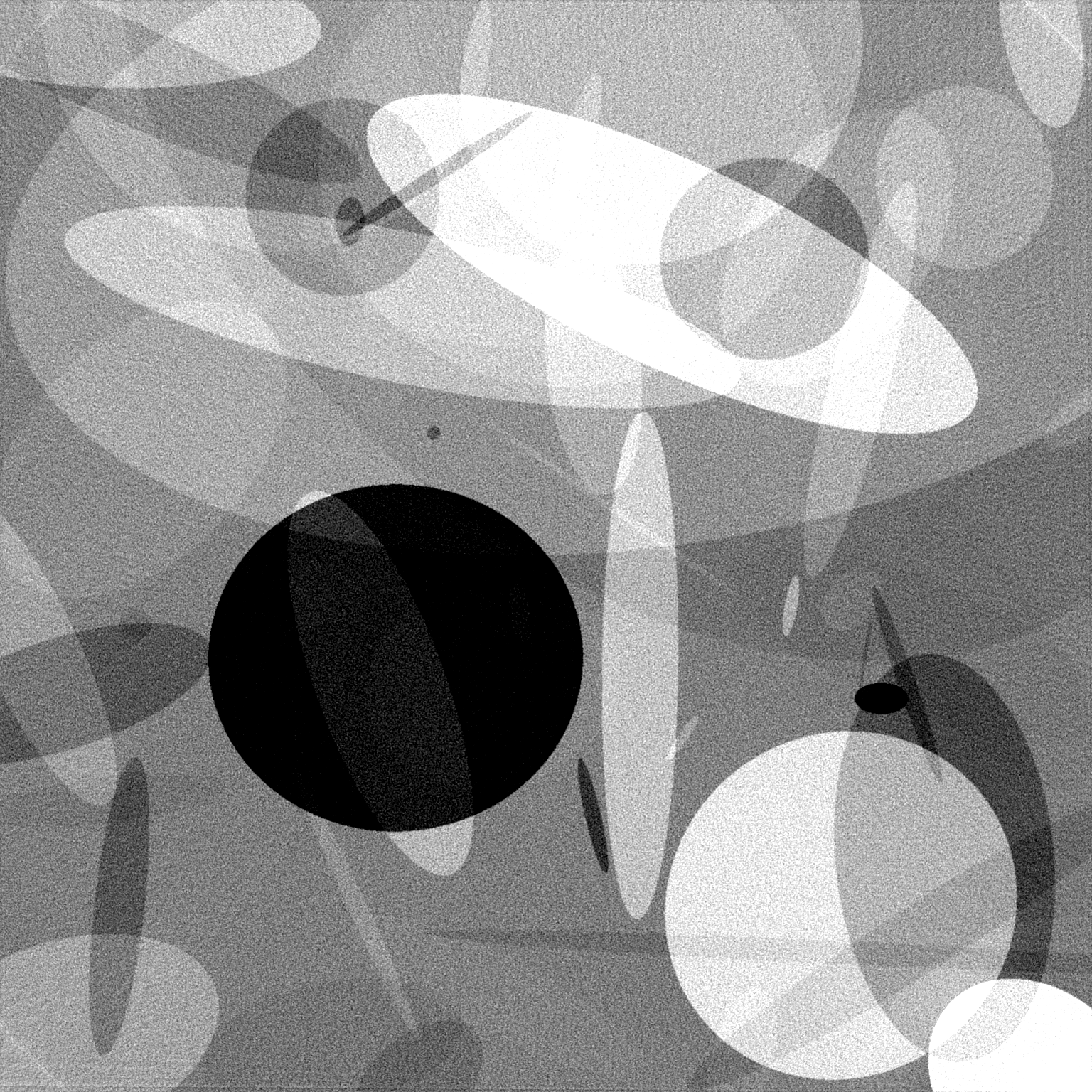}%
\hfill
\showpic{MS-LFGS}{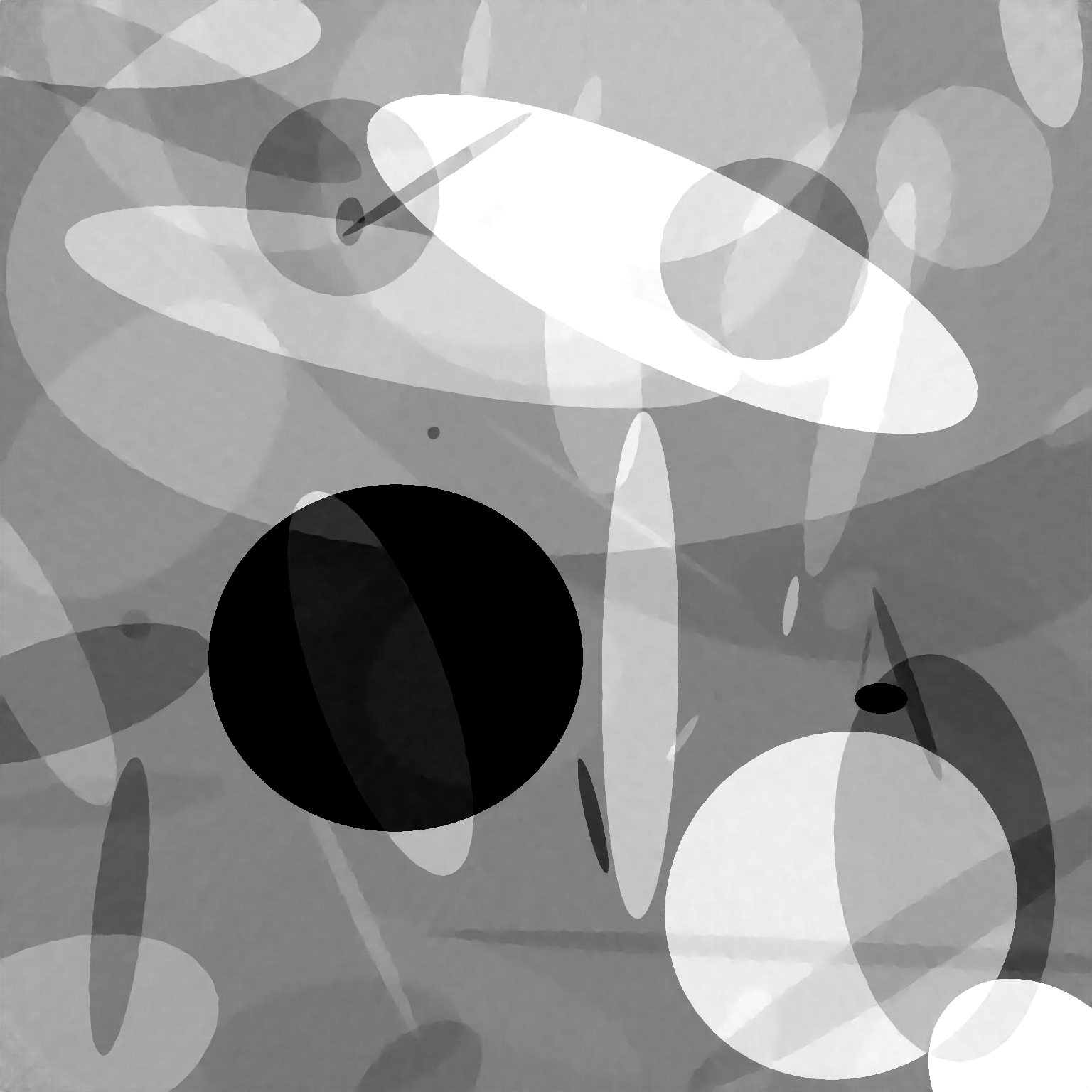}%
\caption{\label{fig:highRes_ellips} Reconstruction of an ellipse phantom of size $1536^2$ from 512 angles with 5\% normally distributed random noise. (Left) Phantom used to create the data, (Middle) reconstruction by filtered backprojection, (Right) obtained reconstruction with \ac{MS-LFGS}.}
\end{figure}}



\subsection{\rev{Memory scaling of reconstruction algorithms}}

Let us first examine the scalability \rev{in terms of memory footprint} of the proposed multi-scale algorithms in comparison to reference learned reconstruction methods. For comparison we choose post-processing with U-Net, following \cite{Jin2017} \rev{with initial channel width of 64, and \ac{LGS} \cite{Adler2017}. Here \ac{LGS}} is implemented consistent with the proposed \ac{MS-LFGS} algorithm, that means we use 5 iterations and a mini U-Net for the sub-networks. In fact, we note that this can be seen as a subclass of \ac{MS-LGS}, where all discretisation spaces are of the same resolution and the scaling operators are given by the identity.

\begin{figure}[t!]
\centering
\begin{picture}(200,165)

\put(-10,5){\includegraphics[width=220pt]{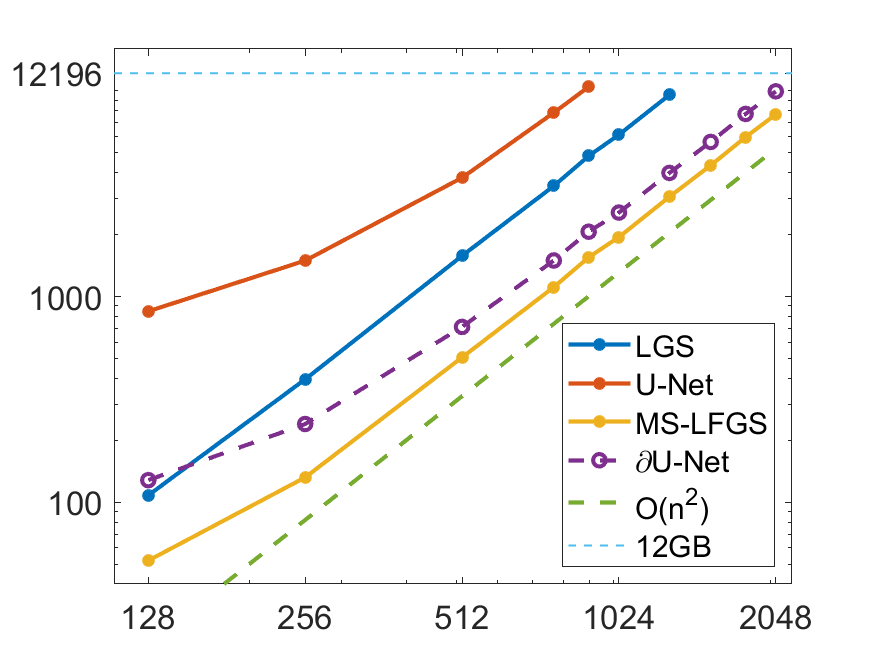}}
\put(65,165){Memory consumption} 
\put(-15,55){\rotatebox{90}{Memory (MB)}}
\put(70,0){Image size: $n\times n$}
\end{picture}
\caption{\label{fig:memory} Memory consumption in the training phase of proposed algorithms and reference learned methods for simulated data in 2D of increasing size. Maximal available memory on the GPU was 12196MB.}
\end{figure}
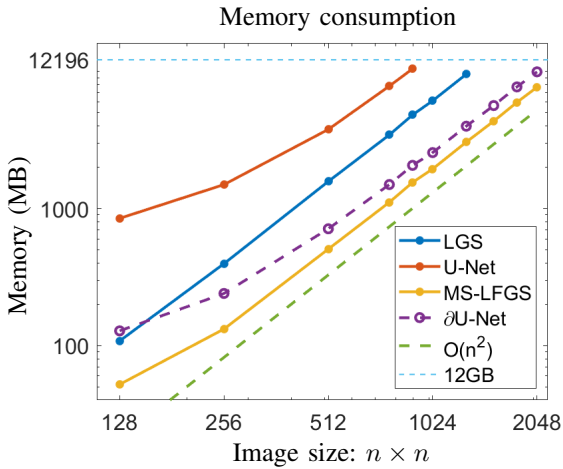



For the training procedure we created phantoms by randomly generated ellipses, see \Cref{fig:highRes_ellips}. The measurement data is then produced by the ray transform \cref{eqn:rayTrafo}, with a fan beam geometry and 512 angles. The simulated measurement is then corrupted by additional $5\%$ of normally distributed random noise.

{\newcommand{\showpic}[2]{%
\begin{tikzpicture}[spy using outlines={circle, magnification=3, 
size=3cm, connect spies}]%
\draw (0,0) node [anchor=south, yshift=-5pt] {\phantom{f}#1\phantom{g}};%
\draw (0,0) node [anchor=north] {\includegraphics[width=3.5cm]{recons/#2}};%
\spy on (-1.25cm,-1.7cm) in node at (-.2cm, -4.75cm);
\end{tikzpicture}\hspace*{-2mm}%
}

\newcommand{\showdata}[2]{%
\begin{tikzpicture}
\draw (0,0) node [anchor=south, yshift=-5pt] {\phantom{f}#1\phantom{g}};%
\draw (0,0) node [anchor=north] {\includegraphics[width=2.65cm]{recons/#2}};%
\end{tikzpicture}\hspace*{-2mm}%
}

\begin{figure}[t!]
\centering
\showpic{Phantom}{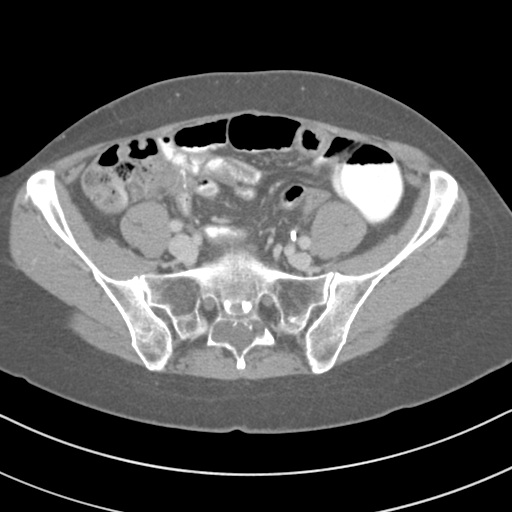}%
\hspace{5mm}
\showdata{Measurement}{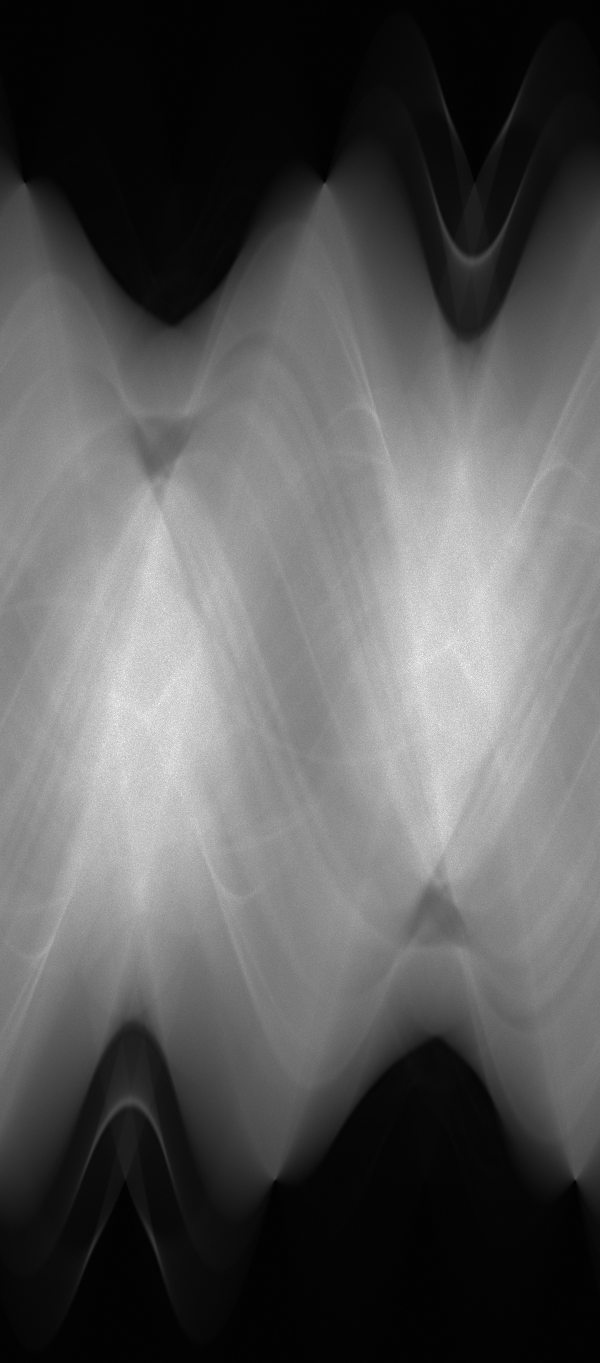}
\caption{\label{fig:phantom} Sample slice from the test patient windowed to $[-300,300]$HU and the corresponding measurement data from 600 angles with \rev{a mean of} 8000 photon counts.}
\end{figure}}

Since the aim of this experiment is to examine \rev{memory consumption only}, we have trained each network for 1000 iterations with one sample in each iteration and recorded the maximum memory consumption. The smallest phantom size was chosen as $128^2$ and was increased until memory consumption exceeded the available memory on a single GPU with 12GB memory, or more specifically 12196 MB. The resulting plot is shown in \Cref{fig:memory}.

\rev{We note that memory consumption of all networks scales with $O(n^d)$, where $d$ is the dimension. A reduction in memory consumption can be mainly achieved by usage of smaller networks and as such reduction by a constant. Nevertheless, memory consumption of \ac{LGS} depends on the number of iterations also, i.e. we have $O(Nn^d)$. Following \Cref{sec:CompCost}, for multi-scale approaches this iteration dependence can be bounded as well by the factor $C_d$ in \eqref{eqn:geometricSeries} and thus we obtain the basic memory dependence of $O(n^d)$.}


A reconstruction obtained with \ac{MS-LFGS} for a resolution of $1536^2$ is shown in \Cref{fig:highRes_ellips} in comparison to a reconstruction by filtered backprojection.



\subsection{Application to human \acs{CT} scans}
In order to evaluate the reconstruction quality on a clinically relevant case, we simulate realistic measurement data from human abdomen \ac{CT} scans provided by the Mayo Clinic for the 2016 AAPM Low Dose CT Grand Challenge \cite{Mccollough2016CT}. The data set consists of high-dose scans from 10 patients. We used the provided reconstructions with 3 mm  slice thickness and image size $512\times 512$. We divided the data into 9 patients for training, resulting in 2168 slices, and 1 patient for testing purposes with 210 slices.

{\newcommand{\showpic}[2]{%
\begin{tikzpicture}[spy using outlines={circle, magnification=3, 
size=2.5cm, connect spies}]%
\draw (0,0) node [anchor=south, yshift=-5pt] {\phantom{f}#1\phantom{g}};%
\draw (0,0) node [anchor=north] {\includegraphics[width=3cm]{recons/#2}};%
\spy on (-1.075cm,-1.475cm) in node at (-.2cm, -4cm);
\end{tikzpicture}\hspace*{-2mm}%
}

\begin{figure*}[th!]
\centering
\showpic{Phantom}{phantom_120_600}%
\showpic{FBP}{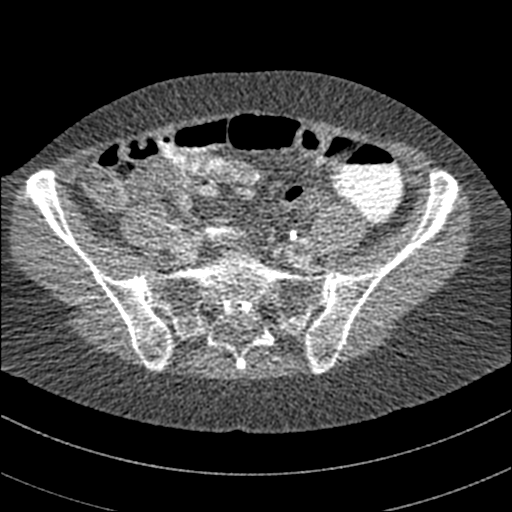}%
\showpic{\rev{U-Net denoiser}}{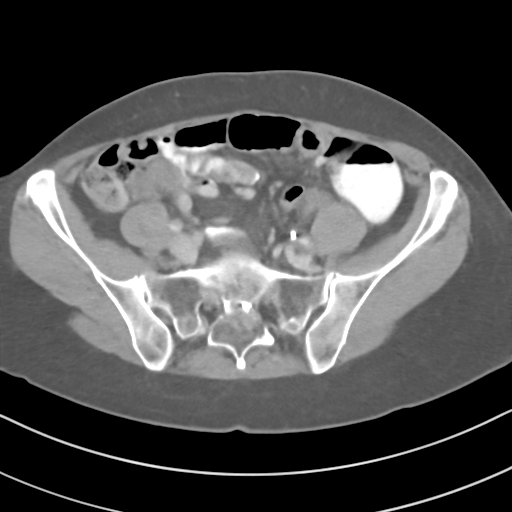}%
\showpic{MS-LFGS}{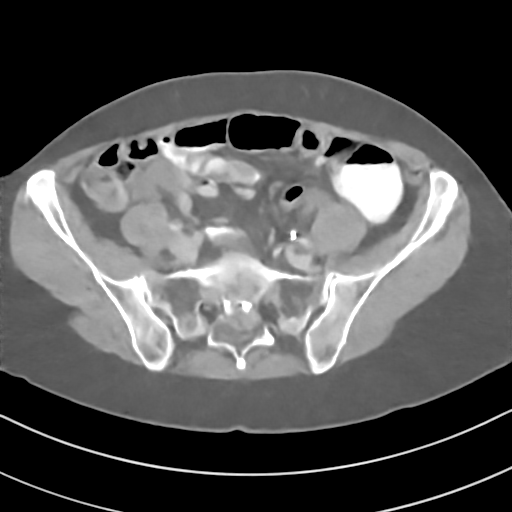}%
\showpic{LGS}{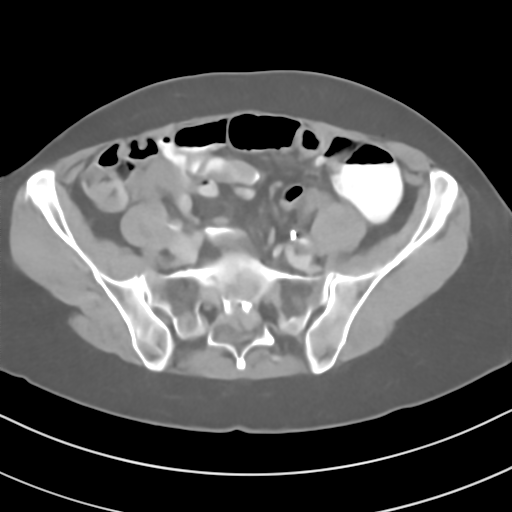}%
\showpic{$\partial$U-Net}{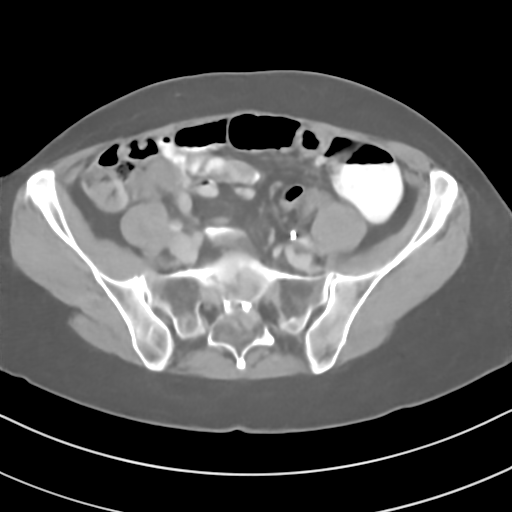}%
\caption{\label{fig:600angleRes} Reconstructions of the test patient for measurement case 1 with 600 angles. All images are windowed and displayed on $[-300,300]$HU. The filtered backprojection here is computed with $h=0.4$.}
\end{figure*}}

For the data simulation, we used a fan beam geometry with source to axis distance 500 mm and
axis to detector distance 500 mm. In order to create realistic measurement data, we use the non-linear forward model given by the Beer-Lamberts law: 
\[
\ForwardOp(\signal)(\ell) = e^{-\mu \int_\ell \signal(x)\, \text{d}x}.
\]
\rev{Here, $\ell$ denotes the line along which the x-ray photons travel and} we select the mass attenuation coefficient $\mu=0.2$ cm$^2$/g, which corresponds approximately to the value of water. We simulate low dose scans \rev{with Poisson noise in the measurement data.}
For the computations we linearise the obtained data by applying $-\log(\cdot)/\mu$ to the measurements, by which the forward model simplifies to the ray-transform as in \cref{eqn:rayTrafo}. A slice from the test patient with the corresponding measurement data is shown in \Cref{fig:phantom}.

We remind that we chose the number of iterations as $N+1=5$ and hence the image resolution in the coarsest discretisation space $S_0$ is just $32\times 32$. For the experiments we consider a scenario that roughly represents a clinical low-dose \ac{CT}  scan with 600 angles and a photon count of 8000.

\subsubsection{Training procedure for low dose scans}
We train both multi-scale schemes as outlined in \Cref{sec:multiScale}, the proposed hybrid $\partial$U-Net, as well as a full-scale learned gradient scheme (LGS) and post-processing with U-Net.
In each case, we compute an initial reconstruction 
by filtered backprojection with the Hann filter and frequency 
scaling of $h=0.6$, this reconstruction is also chosen as the 
input to the post-processing with U-Net.
The same parameters are selected to compute the filtered gradient \cref{eqn:filteredGrad} for the \ac{MS-LFGS}.

To make the comparison uniform for all test cases we trained all algorithms in the same manner. In particular we chose Adam as the optimiser with an $\ell^2$-loss;
each network is trained for 20,000 iterations with one training sample per minimisation step. The initial learning rate is set to $10^{-3}$ with a cosine decay 
These choices have shown to perform well for all presented algorithms.
In the following we will discuss the reconstruction results along with a quantitative evaluation.

\begin{table*}[ht!] 
\small
  \caption{Quantitative measures for low dose scans along with benchmark results for each algorithm. Averaged over 210 slices of test patient. Mean values are shown with their standard deviation.}
  \begin{center}
    \begin{tabular}{l|r|r|r|r|r|r}
&{\sc PSNR} &{\sc SSIM} & {\sc Train} &{\sc Exec.} &{\sc Parameter} &{\sc Memory}    \\
    \midrule
  {\sc \acs{FBP} } &    32.48 $\pm$1.55 &    0.73 $\pm$0.0612 &  $\sim$10s & 33ms & 1 &  1477MB \\ 
  {\sc \acs{LGS} } &   \bf 43.25 $\pm$1.24 &\bf 0.963 $\pm$0.0032 & 2h31m & 149ms & 128970 & 2229MB \\
  \rev{{\sc U-Net denoiser}} &    42.76 $\pm$1.52 &   0.960 $\pm$0.0026 & 1h38m & 67ms & 3.1$\cdot 10^7$ & 2733MB \\
    \hline
  {\sc \acs{MS-LGS}}&   41.42 $\pm$1.33 &    0.948$\pm$0.0041 &  1h42m & 53ms & 128970 & 1143MB \\
  {\sc \acs{MS-LFGS}}&  42.85 $\pm$1.25& 0.960 $\pm$0.0034&  2h07m & 154ms & 129690 & 1143MB \\
  {\sc $\partial$U-Net}& \bf 43.51 $\pm$1.23 & \bf 0.965 $\pm$0.0032& 3h25m & 224ms & 2.3$\cdot 10^6$ & 1351MB \\
    \bottomrule
    \end{tabular}%
  \label{table:lowDoseQuant}%
  \end{center}
\end{table*}%

\subsection{Evaluation of reconstruction quality in 2D}
The resulting reconstructions from 600 angles are shown in \Cref{fig:600angleRes}. Let us first note that U-Net does generally produce sharper images than the learned approaches, but can tend to reconstruct artificial realistic looking features. All iterative approaches tend to produce smoother reconstructions, in particular we observe that in areas of uncertainty the learned approaches are more conservative in recreating features and rather tend to reconstruct uniform areas instead of reproducing features from the training data.


We have computed quantitative measures for all cases as shown in \Cref{table:lowDoseQuant}. Comparing the multi-scale schemes, it is apparent here as well that the filtered gradient is necessary for competitive reconstruction quality. Overall, the proposed $\partial$U-Net does perform best of all algorithms, followed by \ac{LGS} and then \ac{MS-LFGS}. We note here, that it is expected that \ac{LGS} performs better than both multi-scale schemes, as it operates on the full resolution in each iteration, but consequently does not scale very well. Nevertheless, the hybrid network $\partial$U-Net is capable of producing competetive results, while being scalable.

Regarding memory consumption, the multi-scale approaches are expected to be cheapest in terms of memory and training times. Whereas $\partial$U-Net clearly reduces memory consumption in comparison to \ac{LGS}, we can see that here in 2D the training times are slightly longer, due to multiple filtered backprojections in the lower scales. We note that this effect is negated in 3D as seen in \Cref{table:walnutQuant}, since \rev{computational complexity reduces by a factor of 8 on each scale} in 3D instead of just 4 in 2D. It is also interesting to point out that \ac{MS-LGS} is faster in execution times than filtered backprojection followed by U-Net, even though reconstruction quality might not be competitive this can be of use in highly time critical applications.

\section{Discussion}\label{sec:Discussion}
The presented framework for multi-scale learned iterative reconstructions in \Cref{sec:multiScale} provides a general framework for a scalable iterative learned image reconstruction. Combining these multi-scale schemes with a U-Net in the last iterate provides a hybrid network capable of outperforming the previously proposed \ac{LGS} approaches. Nevertheless, as this study is primarily of methodological nature, we would like to discuss in the following a few aspects on how the presented framework can be extended. 

\subsection{The scalability issue}

Recently, some efforts have been made to extend learned iterative reconstruction algorithms to 3D applications. These approaches mainly tackle the memory aspect \rev{of the scalability issue}, which prevents scalability to higher dimensions by hardware restrictions. For instance, by using invertable networks \cite{putzky2019invert} one does not need to store the whole network for computation of the gradient in the training. Whereas this solves \rev{the important issue of memory footprint, it does not address computational complexity of the forward operator and as such is primarily applicable to} forward operators of low complexity, such as the Fast Fourier Transform used in magnetic resonance imaging. For computationally more expensive forward operators, scalability is essentially limited by extensive training times due to the evaluation of the model. The proposed multi-scale schemes provide a possible solution to this dilemma, as the model is only once evaluated on the full resolution. \rev{This is illustrated in \Cref{table:compResources}, where we present the order of computational resources needed for the discussed algorithms in this study. The multi-scale approach addresses both points of the scalability issue, memory footprint and computation times. In comparison to \ac{LGS}, which additionally scales with number of iterations, the multi-scale approach reduces this to the order of a single iteration.}

\begin{table}[h!] 
\small
  \caption{\rev{Scaling properties of discussed algorithms in terms of memory footprint and operator evaluations. Here, $n$ is the image size, $d$ is the image dimension (usually $d=2,3$), and $N$ refers to number of unrolled iterations in learned iterative schemes.}}
  \begin{center}
    \begin{tabular}{l|r|r}
&{\sc Memory} &{\sc Operator eval.}    \\
    \midrule
  {\sc \acs{FBP} } &    $O(n^d)$  & $1$   \\ 
  {\sc U-Net denoiser} &    $O(n^d)$  & $1$   \\ 
  {\sc \acs{LGS} } &    $O(Nn^d)$  & $O(N)$   \\ 
  {\sc Multi-scale } &    $O(n^d)$  & $O(1)$   \\ 
    \bottomrule
    \end{tabular}%
  \label{table:compResources}%
  \end{center}
\end{table}%

In fact, the multi-scale schemes showcase their strength especially in higher dimensions as the reduced evaluation cost scales with the dimension. This can be clearly seen when comparing the study in 2D and 3D as presented here. For instance, the hybrid $\partial$U-Net compared to the basic U-Net has an overhead \rev{in evaluation time} of roughly 300\% in 2D, this reduces to only 50\% in 3D. Which underlines the suitability of the proposed $\partial$U-Net for higher dimensional applications.

\subsection{Influence of scales}
\begin{figure}[t!]
\centering
\begin{picture}(230,400)

\put(0,320){\includegraphics[width=75pt]{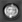}}
\put(0,240){\includegraphics[width=75pt]{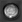}}
\put(0,160){\includegraphics[width=75pt]{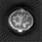}}
\put(0,80){\includegraphics[width=75pt]{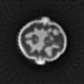}}
\put(0,0){\includegraphics[width=75pt]{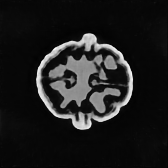}}

\put(80,320){\includegraphics[width=75pt]{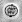}}
\put(80,240){\includegraphics[width=75pt]{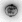}}
\put(80,160){\includegraphics[width=75pt]{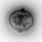}}
\put(80,80){\includegraphics[width=75pt]{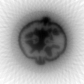}}
\put(80,0){\includegraphics[width=75pt]{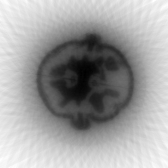}}

\put(160,320){\includegraphics[width=75pt]{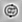}}
\put(160,240){\includegraphics[width=75pt]{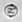}}
\put(160,160){\includegraphics[width=75pt]{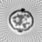}}
\put(160,80){\includegraphics[width=75pt]{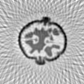}}
\put(160,0){\includegraphics[width=75pt]{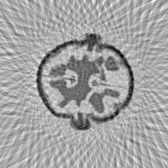}}

\put(7,400){Reconstruction}
\put(97,400){Gradient}
\put(162,400){Filtered Gradient}

\put(-10,340){\rotatebox{90}{Iterate 0}}
\put(-10,263){\rotatebox{90}{Iterate 1}}
\put(-10,180){\rotatebox{90}{Iterate 2}}
\put(-10,100){\rotatebox{90}{Iterate 3}}
\put(-10,20){\rotatebox{90}{Iterate 4}}

\end{picture}
\caption{\label{fig:scaleRecons_600} Representation of the multi-scale scheme used in the $\partial$U-Net with obtained reconstruction, gradient, and filtered gradient on each discretisation space. Reconstructions obtained for the test walnut from 60 angles, here we show the middle slice in the xy-plane.}
\end{figure}
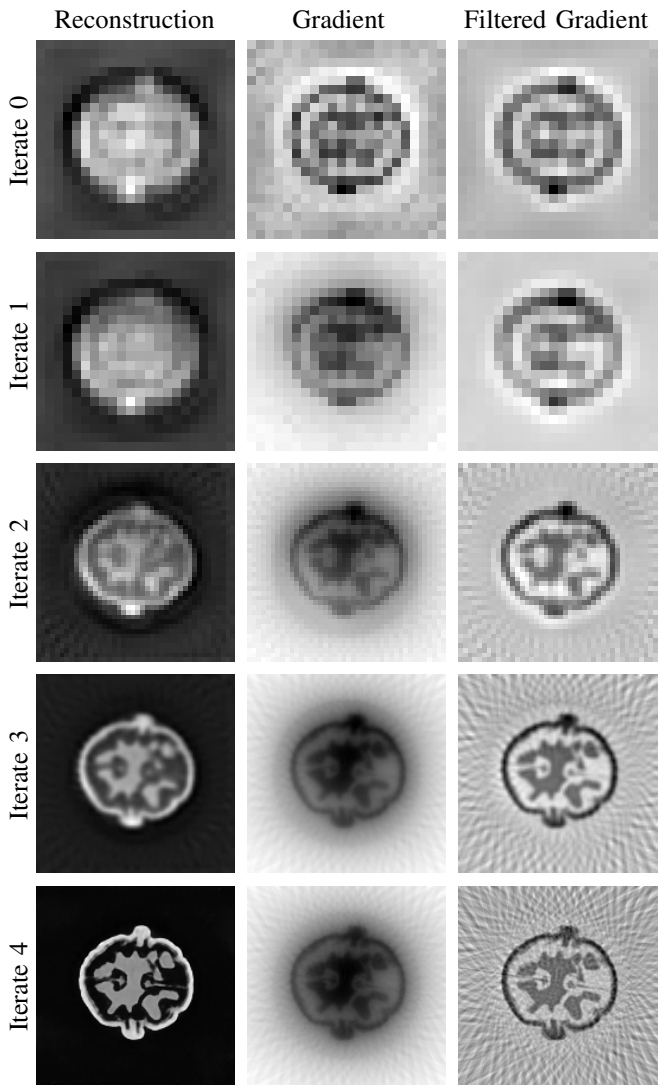

As discussed above, the computational advantage of the multiscale approach is 
primarily due to the low-cost computations on the coarse 
resolution, but these come with some subtleties. We want to note that the early iterates on low resolutions are prone to overfitting and can negatively influence the reconstruction 
quality on the following iterates. Thus, one has to be careful to appropriately deal with the low resolution iterates For instance, the ResNet structure chosen in $\partial$U-Net for the iterative part is more resilient than the mini U-Net. The computed updates in each iterate in the $\partial$U-Net for the walnut reconstructions are shown in \Cref{fig:scaleRecons_600} and as it can be seen the reconstructions nicely gain sharpness in each iterate until the final reconstruction is achieved. 

For the pure multi-scale schemes, as used in MS-LFGS, instead of using only the mini U-Net, one could also consider mixing architectures, especially in the low resolution iterates changing to the ResNet structure for more stability. This has been omitted from this study for the sake of brevity.

\subsection{Extensions of the multi-scale approach}
In this study we have chosen the structure of the multi-scale algorithms as simplistic as possible. Nevertheless, the proposed framework does offer larger flexibility in choices that might be more suitable for other applications. In particular with respect to network design and choice of discretisation spaces. In the following we would like to mention some possibilities how the multi-scale schemes can be extended:
\begin{itemize}
    \item In our study the mini U-Net has shown to be effective to restore high-frequency components more effectively than a basic ResNet style CNN as utilised in \cite{Adler2017}. We note that also more memory efficient networks might be used, such as the MS-D Net \cite{Pelt2018MSD} or, as mentioned above, invertable architectures. Possible extensions of the  $\partial$U-Net to other architectures based on dilated convolutions instead of pooling layers can be investigated as well.
    \item In the multi-scale schemes we have chosen to identify each discretisation space with one iteration. This limitation can be easily relaxed, for instance by computing two iterations in the same discretisation space, as done for the $\partial$U-Net. In case all iterates are computed on the same space, this simplifies to the basic \ac{LGS}. 
    \item We have chosen to reduce the resolution in all dimensions equally. It would be also possible to only reduce the resolution along one dimension in each step and alternate in dimensions. 
    Along the same lines, the up-sampling operator can be chosen differently, including the possibility of a learned up-sampling.
    \item Lastly, the multi-scale framework is not limited to learned gradient schemes and can be extended to other learned approaches such as variational networks \cite{Hammernik2018learning} and learned primal-dual \cite{Adler2018}.
\end{itemize}

\section{Conclusions}\label{sec:Conclusions}
We have presented a general framework for scalable learned iterative reconstruction algorithms for large-scale problems and higher dimensions, by restricting the expensive computation of the \rev{full resolution} forward operator to only one application in the final reconstruction space \rev{and as such reduces computation times as well as memory footprint of the learned iterative scheme.}
This multi-scale approach is especially powerful in higher dimensions, such as 3D, where the computational cost of the early iterates is negligible. 
We have presented two methods to obtain such a scalable learned iterative reconstruction, a basic multi-scale learned (filtered) gradient scheme based on the previous work \cite{Adler2017} as well as hybrid model-based iterative network combined with U-Net, that reuses previously computed gradients on each scale in the respective U-Net scales.

The presented algorithms are evaluated by reconstructing 3D volumes of walnuts from real  measurements, successfully demonstrating scalability of model-based iterative reconstructions to higher dimensions for non-trivial forward operators. The proposed architectures produce competitive results compared to post-processing with U-Net with an increasing robustness due to the iterative model-based nature of the methods. Additionally, we have evaluated the proposed algorithms in 2D in comparison to an established learned gradient scheme, that does not provide easy scalability.

Whereas this work is primarily a methodological study, we believe that it will be of high relevance to applications where high dimensionality of the imaging problem is inherent with computationally demanding forward operators, such as it is the case in cone beam \ac{CT}.

\section*{Acknowledgment}
We acknowledge the support of NVIDIA Corporation with one Titan Xp GPU. The authors also thank Dr. Cynthia McCollough, the Mayo Clinic,  and  the  American  Association  of  Physicists  in  Medicine for providing the data necessary for performing comparison in 2D using a human phantom. 

Codes are available at: \url{https://github.com/asHauptmann/multiscale}
\bibliographystyle{unsrt}
\bibliography{Inverse_problems_references_2018}    
    
\end{document}